\documentclass[10pt,twocolumn,twoside]{IEEEtran}
\usepackage{amsmath,amsfonts}
\usepackage{subfigure}
\usepackage{float}
\usepackage{algorithm}
\usepackage{algorithmicx}
\usepackage{algpseudocode}
\usepackage{mathrsfs}
\usepackage{array}
\usepackage[caption=false,font=normalsize,labelfont=sf,textfont=sf]{subfig}
\usepackage{textcomp}
\usepackage{stfloats}
\usepackage{url}
\usepackage{verbatim}
\usepackage{graphicx}
\usepackage{cite}
\usepackage{bm}
\usepackage{epstopdf}
\usepackage{enumitem}
\newtheorem{lemma}{Lemma}

\begin{document}

\title{	Robust Anti-jamming Communications with DMA-Based Reconfigurable Heterogeneous Array}

\author{
	Kaizhi Huang, Wenyu Jiang, Yajun Chen,  Liang Jin, Qingqing Wu, \IEEEmembership{Senior Member, ~IEEE}, Xiaoling Hu
\thanks{	Kaizhi Huang, Wenyu Jiang,  Yajun Chen and  Liang Jin are with Institute of Information Technology, PLA Information Engineer University, Zhengzhou, China.   Q. Wu is with the Department of Electronic Engineering, Shanghai Jiao Tong University, 200240, China. (e-mail: qingqingwu@sjtu.edu.cn) Xiaoling Hu is with the School of Information and Communication Engineering, Beijing University of Posts and Telecommunications, Beijing, China. (Corresponding author: Wenyu Jiang email: jwy513@hotmail.com) }
}



\maketitle

\begin{abstract}
In the future commercial and military communication systems, anti-jamming remains  a critical issue. Existing homogeneous or heterogeneous arrays with a limited degrees of freedom (DoF) and high consumption are unable to meet the requirements of communication in rapidly changing and intense jamming environments. To address these challenges, we propose a reconfigurable heterogeneous array (RHA) architecture based on dynamic metasurface antenna (DMA), which will increase the DoF and further improve anti-jamming capabilities. We propose a two-step anti-jamming scheme based on RHA, where the multipaths are estimated by an atomic norm minimization (ANM) based scheme, and then the received signal-to-interference-plus-noise ratio (SINR) is maximized by jointly designing the phase shift of each DMA element and the weights of the array elements. To solve the challenging non-convex discrete fractional problem along with the estimation error in the direction of arrival (DoA) and channel state information (CSI), we propose a robust alternative algorithm based on the S-procedure to solve the lower-bound SINR maximization problem. Simulation results demonstrate that the proposed RHA architecture and corresponding schemes have superior performance in terms of jamming immunity and robustness.
\end{abstract}


\section{Introduction}

\IEEEPARstart{A}{nti-jamming} technology is a crucial aspect of modern communication systems that aims to prevent or mitigate intentional or unintentional interference with signal transmission and reception. Due to the openness of the wireless environment, the receivers are susceptible to jamming by signals from other communication systems. Especially that the emergence of 5th Generation (5G) networks which promise to deliver ultra-fast, low-latency, and high-reliability services for various applications, the jamming attacks have introduced new threats and vulnerabilities to the entire  system. Therefore, anti-jamming technology plays a vital role in enhancing the security and robustness of wireless communication networks.

Traditional anti-jamming techniques focused on the time and frequency domains, and then developed a series of schemes by exploiting the signal characteristics in the time and frequency domains, such as frequency hopping (FH) and direct sequence spread spectrum (DSSS) \cite{SUR2}. With the development of multi-antenna and digital signal processing technology, spatial anti-jamming has emerged as a new approach. Spatial anti-jamming adjusts the complex gain of each array element in the antenna array to steer a main lobe towards the desired signal source and nulls towards the undesired sources\cite{AB2,AB1}. In general, spatial anti-jamming schemes predominantly use homogeneous arrays, which have advantages such as ease of processing and the ability to obtain comprehensive information. However, the number of array elements must continuously increase to effectively improve anti-jamming capability \cite{AB3}, leading to high system complexity and hardware costs. It is also a challenge for large-scale deployment due to the antenna array density constraints imposed by the back-end RF link. Moreover, homogeneous arrays are susceptible to the homo-morphic issue. Since the signals are superimposed in the same way for all directions, the  deep fading caused by the inverse multipath, and signal blocking problem(the signal is lower than  the analog to digital converter's (ADC)  Least Significant Bit (LSB)  \cite{ADC1})  can spread to all arrays and significantly affect array performance. In summary, there are limitations to the current homogeneous arrays in terms of adaptive anti-jamming.

Therefore, there has been a growing interest in heterogeneous antennas to improve  array resolution and resist jamming. Heterogeneous antennas consist of an antenna system that can vary antenna radiation patterns on demand. The heterogeneous antennas can be classified into physical-change and electronic-change schemes. However, physical-change schemes are limited by the change speed and the controllability, while the cost and power consumption of electronic-change schemes will increase significantly. Metamaterials bring new opportunities for enhancing anti-jamming capabilities, which has become a potential technology for future 6th generation (6G) systems. The special electromagnetic characteristics and the simple structure allow them to be densely deployed  at $\lambda/5$ to $\lambda/10$ spacings \cite{Hwang2020}, further enabling sub-wavelength level control of electromagnetic waves with lower power and cost consumption. What's more, metamaterials are capable of adjusting the state of the element within the nanosecond through FPGA \cite{MM1}.

As a result, metamaterials are rapidly promoted in variety of applications. On the one hand, metamaterials are deployed in the transmission propagation and serve as intelligent reflecting surfaces (IRS) to provide auxiliary functions for multi-user communication and anti-jamming\cite{RIS1,RIS2,RIS3,RIS4}. On the other hand, dynamic meta-surface antennas (DMA) use metamaterials as the basic elements of the antenna\cite{DMA2}. Compared to traditional phased or digital arrays, DMA-based architectures require less energy and cost consumption while transmitting and receiving signals with simplified hardware and dynamic configuration. Through dense deployment, DMA can provide almost arbitrary aperture fields  and outperform in terms of degrees of freedom (DoF), estimation accuracy and control speed\cite{DMA1}. In this paper, we focus on enhancing the anti-jamming capability of the receiver using DMA with their dense deployment and fast switching ability.

\subsection{Related works and Motivation}

 Numerous heterogeneous antennas-based schemes have been proposed for anti-jamming. One way to implement heterogeneous antennas is through physically reconfigurable antenna technology. For instance, the authors in \cite{ESPAR1} used an parasitic element  to generate adjustable radiation patterns. Without changing the antenna itself, Wang et al. formed patterns with different features by connecting a pixel antenna array differently\cite{Rad1}. Another approach is beam switching, which selects different beams at different times to generate the time-division heterogeneous antenna\cite{RA2}. While physical heterogeneous antennas can increase DoF and enhance anti-jamming capabilities in specific directions, it is challenging to model their radiation patterns directly. Deploying different antennas will sacrifice the flexible processing capability of homogeneous arrays. Additionally, the mechanical or switching operations is hard to catch up with the signal sampling.

As for the electronic-change schemes, researchers have explored heterogeneous arrays with multiple subarrays to improve the DoF of the system. In radar systems, Passive Electronically Scanned Arrays (PESA) with multiple phased array units exploit the interference effect of received signals. Specifically, the authors in \cite{PHA3} proposed an algorithm to find the optimal subarray configuration in large-scale phased arrays, and used adaptive beamforming based on partitioned subarrays to resist jamming. To improve the anti-jamming capability, the authors in \cite{PHA2} designed the subarray structure and the beamforming vector of the transmitter to improve the anti-jamming ability of the phased-MIMO.  For wireless communications, the hybrid beamforming (HBF) technique combines passive antennas with a few digital RF channels, which take advantage of digital processing with analog phase shifts. The authors in \cite{HB3} proposed an HBF architecture to reduce hardware and power consumption. Hybrid beamforming vectors are jointly optimized by a heuristic algorithm and achieve full RF link performance with few RF links. Wen et al. \cite{HB1} further considered the finite resolution phase shifters and designed the digital and analog beamforming vectors to improve communication rates in multi-user systems. However, perfect channel state information (CSI) is difficult to obtain, especially in a complex jamming environment. The authors in \cite{HB2} designed a robust HBF scheme in the multi-user multi-cell mmWave system.  Although schemes such as HBF have reduced the number of RF links, they are still challenging to  deploy densely due to  the large number of physical components such as high-frequency and high precision phase shifters.

Based on the low cost and massive-elements characteristics, DMA can effectively improve the antenna's DoF and improve the aperture efficiency, which is quickly applied in sensing \cite{DMAS3} and communications\cite{DMAC2,DMAC3,DMAC4,DMAC5,DMAC6,DMAC7}. Specifically, in order to improve the accuracy of direction of arrival (DoA) estimation, the authors in \cite{DMAS3} proposed a spatial smoothing scheme that can provide higher resolution compared to homogeneous arrays. Shlezinger et al. in \cite{DMAC2} first applied DMA as a large-scale transmit-receive array for future 6G, and analyzed the differences between DMA and traditional HBF methods. In \cite{DMAC7}, the authors deployed DMA at the transmitter and receiver simultaneously in MU-MIMO networks, and an alternating optimization algorithm was proposed to improve the uplink and downlink rates. In MIMO-OFDM systems \cite{DMAC6}, DMA is used to reduce the cost and power consumption. By establishing the relationship between frequency and DMA response, a spectrally flexible hybrid structure is constructed. In addition, the paper considered operating with bit-constrained ADCs to facilitate recovering transmitted OFDM signals using low-resolution ADCs. The works in \cite{DMAC4} and \cite{DMAC5} further improved the communication performance of large-scale MIMO networks by combining the advantages of DMA and IRS.

Although the DoA estimation and the beamforming design of DMA for MISO, MIMO and near-field communication have been investigated, few studies have been conducted to construct heterogeneous antenna with multiple DMA and improve the estimation accuracy and anti-jamming capability.  Since the weights of each element in DMA are constrained by the electromagnetic structure, traditional schemes are hardly applied to DMA. Moreover, the actual estimation results contain a certain amount of error, which has a large impact on the anti-jamming  performance.  The estimation uncertainty and the phase shift constraints will further challenge the anti-jamming design which have not been considered in the above mentioned heterogeneous antenna-based anti-jamming schemes. Therefore, it is crucial to study the channel estimation and further robust anti-jamming scheme design with the heterogeneous arrays.

\subsection{Contributions and Organization}

In light of the significant advantages of DMA, we exploit DMA to construct a reconfigurable heterogeneous array (RHA). In order to eliminate strong jamming, we first estimate the DoA and CSI of signals and jamming through the fast switching ability of RHA. With the help of RHA, we try to superimpose the signal multipaths in-phase, while the jamming multipaths are mutually weakening. Furthermore, considering the estimation error, a robust anti-jamming scheme is formulated for maximizing the lower bound of the received SINR. To the best of the author's knowledge, this is the first work that uses RHA to improve the anti-jamming ability.

In summary, the contributions of this paper are as follows:
\begin{enumerate}

	\item We present the model of the finite scattering environment, and compare the generalized receiving model for the homogeneous arrays and heterogeneous array.  Since the ideal heterogeneous antenna does not exist, we propose a novel RHA based on the ultra-dense deployment and pattern-changing ability of the DMA, and give the signal model of the DMA-based RHA.

	\item We propose a two-step anti-jamming scheme based on the RHA model. Firstly, the multi-path DoA and CSI are estimated by an off-grid atomic norm minimization-based (ANM)  scheme. Secondly,  to deal with the uncertainty of  the DoA and CSI estimation errors, the bounded estimation errors are considered, and the robust anti-jamming  scheme is finally transformed into a minimum SINR maximization problem.
	
	\item  The optimization problem is challenging due to non-convex fractional problems and the infinity constraints caused by the estimation uncertainty. Thence, an alternating optimization algorithm based on the S-procedure and discrete approximation method are proposed  to solve the problem of maximizing SINR lower bound. 
	
	\item Simulation results show that the DMA-based RHA has advantages in resisting strong multipath jamming and improving energy efficiency under the finite scattering environment. What's more , the results verify the effectiveness of the proposed anti-jamming scheme in terms of robustness compared to other schemes under various jamming powers.

\end{enumerate}

The rest of the paper is organized as follows. In Section II, the model of the ideal heterogeneous array and the realistic model of DMA-based RHA are given.  In Section III, an anti-jamming scheme is proposed which consists of the ANM-based estimation algorithm and the robust scheme for RHA with estimation errors. Simulation results are given in Section IV. Finally, the paper is concluded in Section V.

{\emph{Notations: }} Boldface lowercase and uppercase letters denote vectors and matrices, respectively. ${\bf v}=[v_n],n=1,\cdots,N$ represents the vector ${\bf v}$ consists of values represented by different subscripts as $[v_1, \cdots, v_N]$. The transpose, conjugate transpose, rank, and trace of the matrix $A$ are denoted as ${\bf{A}} ^T$, ${\bf{A}} ^H$, Rank(${\bf{A}} $), and Tr(${\bf{A}} $), respectively. ${\bf{A}} {\succeq} 0$ means ${\bf{A}}$ is a positive-semidefinite (PSD) matrix. $\left\|{\cdot{}}\right\|$ denotes the vector Euclidean norm. ${\bf{A}}\otimes{\bf{B}}$  and ${\bf{A}}\odot{\bf{B}}$ represent the Kronecker  and Hadamard product of matrices ${\bf{A}}$ and ${\bf{B}}$, respectively. ${\cal C}{\cal N}(\mu,\sigma^2)$ denotes a complex circularly symmetric Gaussian distribution with a mean value $\mu$ and variance $\sigma^2$. $\mathbb{E}(x),\mathbb{D}(x)$ represent the expectation and variance of the random variables, respectively. ${\bf I}$  is the identity matrix.

\section{System Model}

\begin{figure}[t]
	\begin{center}
		\includegraphics[scale=0.5]{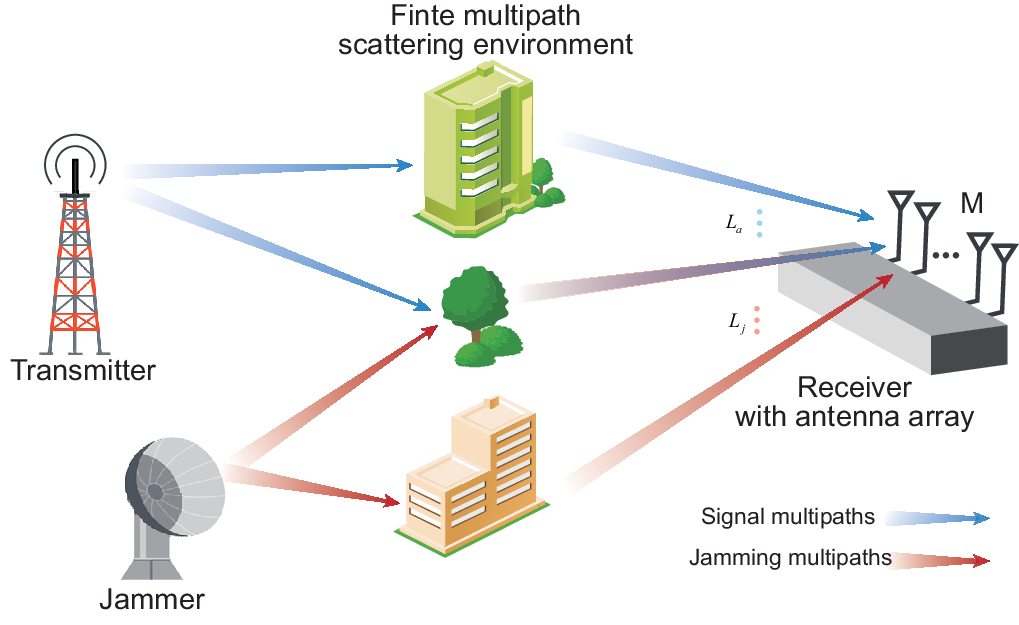}
		\caption{The illustration of the system model under finite multipath scattering environment }\label{fig0}
	\end{center}
\vspace{-0.5cm}
\end{figure}

We consider a finite scattering environment in the presence of a jammer, which is illustrated in Fig.\ref{fig0}. The legitimate transmitter (denoted as Alice) sends information to the receiver through an omni-directional antenna, while the corresponding receiver Bob tries to receive Alice's information through an antenna array and avoids jammer's influence.

The Salen-Valenzula model is used to model multipaths. We assume that the numbers of incoming multipaths of Jammer (denoted as $L_j$) and Alice (denoted as $L_a$) are finite. The DoA of the jamming signals are denoted as ${\boldsymbol{\theta }_j}=[\theta_{j,1},\cdots,\theta_{j,L_j}]$ where $L_j$ is the number of the multipaths, and the DoA of the desired signals are denoted as ${\boldsymbol{\theta }_a}=[\theta_{a,1},\cdots,\theta_{a,L_a}]$ where $L_a$ is the number of the multipaths. Each multipath channel obeys a Circle Complex Gaussian distribution in a coherence time, which is denoted by
\begin{equation}\label{eq1}
	{\bf{g}}_{zb}^{} = \left[ g_{zb,1}, \cdots, g_{zb,L_z}\right] \in {\mathbb{C}^{{L_z} \times 1}},
\end{equation}
where $g_{zb,l}={\mu _z^{\left( l \right)}{e^{j\tau _z^{\left( l \right)}}}}$ and ${\mu _z^{\left( l \right)}}$ and $ {\tau _z^{\left( l \right)}}$ are the  path loss and delay of the $l$-th multipath, respectively. For simplicity, we denote $z=a$ or $j$ to indicate that the parameter belongs to Alice or Jammer, which is not repeated below.

In order to improve the anti-jamming ability of the receiver, Bob uses an antenna array with $M$ elements to receive the signals coming from multipaths. Based on the generalized signal reception model,  the electromagnetic (EM) wave is mapped by the antenna into a mixed signal. The received signal at the $i$-th antenna can be expressed as
\begin{equation}\label{eq2}
	\begin{aligned}
	{y_i} =& \sum\limits_{z =s,j} {\left( {\boldsymbol{\phi }_i({{\boldsymbol{\theta }}_z}) \odot {{\bf{a}}_i}({{\boldsymbol{\theta }}_z})} \right)^H}{{\bf{g}}_{ab}}{s_a} + {n_i},
\end{aligned}
\end{equation}
where ${\boldsymbol{\phi} _i}({\boldsymbol{\theta }_z}) = {[\alpha ({\theta _l}){e^{j\beta ({\theta _l})}}]^T},l = 1 \cdots {L_z}$ denotes the complex gain of the antenna for different angles, which includes the amplitude response $\alpha ({\theta _l})$ and phase response $e^{j\beta ({\theta _l})}$. In the traditional uniform line array (ULA), the receiving antenna pattern is generally isomorphic and omni-directional. ${{\bf{a}}_i}({{\boldsymbol{\theta }}_z}) = {[{e^{ - j k_{0}(i - 1)d \sin ({\boldsymbol{\theta }_z})}}]} \in \mathbb{C}^{{L_z} \times 1}$ denotes the phase deflection of the $i$-th antenna to the first antenna, $k_{0}=2\pi/\lambda$ is the wave number, and $d$ is the spacing between two antennas. ${{\boldsymbol{\theta }}_z}$ and ${s_z}$ denote the DoA and symbols, respectively. ${n_i} \sim {\cal CN}(0,\sigma_n^2)$ is the complex additive white Gaussian noise (AWGN).

The signal is first amplified by a low-noise amplifier (LNA). The receiver then converts the analog signal to a digital signal through a uniform quantization ADC and performs subsequent digital signal processing. Taking IQ sampling as an example, the mixed signal after the ADC is further transformed into
\begin{equation}\label{eq3}
	{\hat y_i} = Q(A_i{y_i}),
\end{equation}
where $A_i$ is the gain of the LNA, $Q(\bullet)$ indicates the sampling and quantization operations performed by the ADC. The LSB of a K-bits ADC can be defined as $LS{B} = {V_{f}}/{2^K}$ , where  ${V_{f}} $ is the full-scale range (FSR).  When the jamming environment is extremely severe, it will push the amplifier to the nonlinear region, which leads to saturation distortion of the signals. What's more, when the signal-to-jamming ratio is too small, the ADC blocking occurs.  For instance, if there are two signals denoted as $a_r+ja_i$ and $b_r+jb_i$ enter the ADC, if $b_r/(a_r+b_r)$ or $b_i/(a_i+b_i)$ is smaller than LSB, the information of $b$ will be modified or even missing. Without loss of generality, we assume that the LNA always operates in the linear region with $A_i = 1$ and path gain is below the full scale range of the ADC. The effect of $Q$ can be neglected if ADC blocking does not occur. The SINR of the receiver is
\begin{equation}\label{eq3a}
	{\rm{SINR}}	= \frac{{P_a\vert\sum\limits_{i = 1}^M {\rm{w}}_iQ({A_i{{\left( {{\phi _i}({{\boldsymbol{\theta }}_a}) \odot {{\bf{a}}_i}({{\boldsymbol{\theta }}_a})} \right)}^H}{{\bf{g}}_{ab}}})\vert^2 }}{{P_j\vert\sum\limits_{i = 1}^M {\rm{w}}_iQ({A_i{{\left( {{\phi _i}({{\boldsymbol{\theta }}_j}) \odot {{\bf{a}}_i}({{\boldsymbol{\theta }}_j})} \right)}^H}{{\bf{g}}_{jb}}})\vert{^2}  + {\sigma_n ^2}}},
\end{equation}
where ${\bf{w}} = [{\rm{w}}_i],i = 1 \cdots M$ is the weights of each antenna.

\subsection{Receiving model for Homogeneous Array}

Considering homogeneous arrays which is illustrated in Fig.\ref{fig1}, the gain in different angle are ${\boldsymbol{\phi} _i}({\boldsymbol{\theta }_z}) = {[1,\cdots,1]}$. The channel after the antenna sampling can be simplified as ${h_{i,zb}} = {{\bf{a}}_i}({{\boldsymbol{\theta }}_z})^H{{\bf{g}}_{zb}}$. In such cases, the homogeneous arrays can estimate the DoA of jamming and signal, and obtain the optimal beamforming \cite{Wang2021a}. 

But the homogeneous arrays have certain limitations. On the one hand, the adjustment of the phase in the homogeneous array can only be performed after the antenna. The multi-path signals have been superimposed during the antenna reception, and the equivalent channel of one antenna exhibits small-scale fading.  Therefore, the communication quality can only be improved by the array gain with a large number of antennas.

However, if ${{\bf{a}}_i}({{\boldsymbol{\theta }}_z})^H{{\bf{g}}_{zb}}=0$, such as ${\theta _1} = -{\theta _2}=arcsin(\lambda/4d)$ and $g_1=g_2$, the signals will be inverse-superimposed thus producing severe fading no matter how the weights on each antenna are adjusted.  On the other hand, the jamming signals are mixed arbitrarily on the omni-directional antenna, resulting in a high probability of ADC blocking.  It is no longer possible to separate  the legitimate signal from the mixed signal at the back-end, and the digital signal processing schemes will fail.  What's more, the coherent multipaths will make the subspace of the signal and jamming not full-rank. The spatial resolution of the homogeneous array is limited, which reduces the accuracy of estimation and anti-jamming performance. 


\subsection{DMA-based Reconfigurable Heterogeneous Arrays}

To address these issues, we propose an ideal heterogeneous array. In brief, if we can combine multi-path signals as desired during the antenna reception instead of disorderly superposition, it can match the multipath signals reduce the aforementioned problems. Unlike the homogeneous array's element, each antenna in the heterogeneous array can control its complex gain in different incoming directions arbitrarily\cite{Guo2018}. According to (\ref{eq1}),  we can design a radiation  pattern for $i$-th antenna which satisfies
\begin{subequations}\label{eq4}
	\begin{gather}
		\begin{aligned}
			&\boldsymbol{\phi}_i {({{\boldsymbol{\theta }}_a})^H}{{\bf{g}}_{ab}} = \sum\limits_{l = 1}^{{L_a}} {\phi_i ({\theta _{a,l}}){g_{ab,l}}}  = \sum\limits_{l = 1}^{{L_a}} {|\phi_i ({\theta _{a,l}})||{g_{ab,l}}|},\\
			&\qquad\qquad\qquad\qquad |\boldsymbol{\phi}_i {({{\boldsymbol{\theta }}_j})^H}{{\bf{g}}_{jb}}{|^2} = 0.
		\end{aligned}
	\end{gather}
\end{subequations}
where (\ref{eq4}a) makes the signal multipaths in-phase and makes the gain as large as possible in every direction. (\ref{eq4}b) ensures that the pattern is orthogonal to the jamming channel without constraint on the gain. It should be noted that although we do not set the gain in each direction, the maximum gain is still limited by the size of the antenna.

\begin{figure}[t]
	\begin{center}
		\includegraphics[scale=0.9]{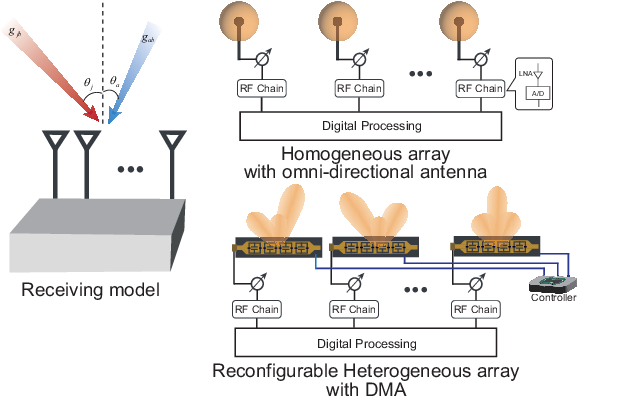}
		\caption{The illustration of receiving model of the homogeneous array with omni-directional antenna and RHA. The RF link consists of an LNA and an ADC, then the signal is handled by digital processing. For RHA, each DMA's phase shifts are controlled by  a controller to perform various radiation patterns. }\label{fig1}
	\end{center}
	\vspace{-0.5cm}
\end{figure}

Although the above ideal heterogeneous array brings us a significant advantage, it is inaccessible in reality. In order to approximate the capability of the heterogeneous array, we propose a DMA-based RHA array. According to the existing meta-surface structures \cite{Hwang2020}, the spacing between elements (noted as $d_e$) can be generally lower than $\lambda/2$. Thanks to the low cost and simple structure of meta-surface elements, we can deploy a large amount elements in a limited area, which provides a good basis for the implementation of heterogeneous antennas, as shown in Fig.\ref{fig1}.  It is assumed that there exists a heterogeneous array of $M$ heterogeneous antennas in the horizontal direction with spacing $d$.  Each heterogeneous antenna is constructed by the DMA and contains $N$ meta-material elements which are located on a waveguide with spacing $d_e$. The elements are able to adjust the phase of the EM wave through the internal components (such as variable capacitors of pin-diodes). The radiation pattern of the $m$-th DMA  is
\begin{equation}\label{eq13}
	{\boldsymbol{\hat \phi} _m}({\boldsymbol{\theta }}_z)  = {\boldsymbol{\omega }}_m^H{{\bf{T}}}{\boldsymbol{\delta }}({\boldsymbol{\theta }}_z).
\end{equation}
where ${{\boldsymbol{\omega }}_m} = {[{\omega _{m,1}}, \cdots {\omega _{m,N}}]^H} \in {\mathbb{C}^{N \times 1}}$ represents the phase shifts of the $m$-th DMA elements. ${{\bf{T}}_m}=diag([{t_{m,1}},\cdots,{t_{m,N}}]) \in {\mathbb{C}^{N \times N}}$ is a diagonal matrix represented the transmission attenuation on the waveguide with ${t_{m,n}} = {e^{ - {r_n}({\alpha _t} + j{\beta _t})}}$, where ${\alpha _t}$ and ${\beta _t}$ are the attenuation and  propagation factors.  $r_n=(n-1)d_e$ is the location of the $n$-th element. The DMA manifold vector ${\boldsymbol{\delta }}({\boldsymbol{\theta }}_z) $ is defined as
\begin{equation}\label{eq14}
	{\boldsymbol{\delta }}({{\boldsymbol{\theta }}_z}) = \left[ {\begin{array}{*{20}{c}}
			{{e^{ - jk_0 {r_0}\sin {\theta _{z,1}} }}}& \cdots &{{e^{ - jk_0 {r_{0}}\sin {\theta _{{z,L_z}}} }}}\\
			\vdots & \ddots & \vdots \\
			{{e^{ - jk_0 {r_{N-1}}\sin {\theta _{z,1}} }}}& \cdots &{{e^{ - jk_0 {r_{N-1}}\sin {\theta _{{z,L_z}}} }}}
	\end{array}} \right].
\end{equation}

Although the elements in the existing DMA can only perform several states, an approximation to RHA can be achieved by large-scale and dense deployment of meta-elements. We consider a commonly used pin-diode based meta-materials, and define the set of the available phase shifts as ${\boldsymbol \Psi} = \left\{ {{{2\pi b}}/{{{2^B}}}} \right\},b = 0, \cdots ,{2^B} - 1$, where $B$ is the control bits. The received signal of the RHA in matrix form is
\begin{equation}\label{eq15}
	{\bf{y}} = {{{\bf{w}}^H}{\boldsymbol{\Phi }}({{\boldsymbol{\theta }}_a}) \odot {\bf{A}}({{\boldsymbol{\theta }}_a}){{\bf{g}}_{ab}}}s_a+{{{\bf{w}}^H}{\boldsymbol{\Phi }}({{\boldsymbol{\theta }}_j}) \odot {\bf{A}}({{\boldsymbol{\theta }}_j}){{\bf{g}}_{jb}}}s_j+{\bf{n}}.
\end{equation}
where  ${\boldsymbol{\Phi }}({{\boldsymbol{\theta }}_z}) = {[\boldsymbol{ \hat \phi }_1({{\boldsymbol{\theta }}_z})^T,\cdots, \boldsymbol{ \hat \phi }_M({{\boldsymbol{\theta }}_z})^T]^T} \in \mathbb{C}^{M \times L_z}$ is the pattern matrix , and ${\bf{A}}({{\boldsymbol{\theta }}_z}) = {[{\bf{a}}_1({{\boldsymbol{\theta }}_z}),\cdots,{\bf{a}}_M({{\boldsymbol{\theta }}_z})]}^T \in \mathbb{C}^{M \times L_z}$ is the array steering matrix. The received SINR of RHA is further simplified by
\begin{equation}\label{eq16}
	SINR = \frac{{{P_a}|{{\bf{w}}^H}{\boldsymbol{ \Phi }}({{\boldsymbol{\theta }}_a}) \odot {\bf{A}}({{\boldsymbol{\theta }}_a}){{\bf{g}}_{ab}}{|^2}}}{{{P_j}|{{\bf{w}}^H}{\boldsymbol{\Phi }}({{\boldsymbol{\theta }}_j}) \odot {\bf{A}}({{\boldsymbol{\theta }}_j}){{\bf{g}}_{jb}}{|^2} + \sigma _n^2}}.
\end{equation}

\section{Anti-jamming Scheme based on RHA }
In this section, a two-stage anti-jamming scheme based on RHA are proposed to improve the SINR.  In the first estimation process, the DoA and CSI of the signal and jamming are estimated. In the subsequent anti-jamming process, the maximum anti-jamming performance is achieved by jointly designing the weights of antennas and the phase shifts of DMA elements.

\subsection{DoA and CSI Estimation based on RHA}

It is crucial to obtain the DoA and CSI of the jamming and the signal for the subsequent design. Therefore, we propose a RHA-based high-precision DoA and CSI estimation method to estimate the DoA by an off-grid method based on ANM and then obtain the CSI of each multipath accordingly. Without loss of generality, the pilot symbols are denoted as ${{\bf{x}}_z} = [{x_{z,1}}, \cdots ,{x_{z,T}}]^T$. To achieve high accuracy estimation of DoA while mitigating the impact of ADC blocking, we propose a spatial-temporal DoA estimation through the rapid pattern-changing ability and the array structure of RHA.

We sample $K_r$ times with different radiation patterns, and the sample points all fall in the period where the symbols  have not changed. The $k$-th pattern matrix of the DMA is defined as ${{\bf{\Phi }}_k}({{\bf{\theta }}_z})$. The received signal is expressed as
\begin{equation}\label{eq17}
	{{\bf{Y}}_{k,t}} = {{\boldsymbol{\Phi }}_k}({{\boldsymbol{\theta }}_a}) \odot {\bf{A}}({{\boldsymbol{\theta }}_a}){{\bf{g}}_{ab}}{x_{a,t}} + {{\boldsymbol{\Phi }}_k}({{\boldsymbol{\theta }}_j}) \odot {\bf{A}}({{\boldsymbol{\theta }}_j}){{\bf{g}}_{jb}}{x_{j,t}} + {{\bf{n}}_k}.
\end{equation}

For dealing with the form of the Hadamard product, we introduce the following lemma:
\begin{lemma}\label{lem2}
	The Hadamard product of two matrices ${\bf A}=[{{\bf a}_1},\cdots,{{\bf a}}_M]^T$ and ${\bf B}=[{{\bf b}_1},\cdots,{{\bf b}}_M]^T \in \mathbb{C}^{M \times N}$ times a vector $\bf w$ $\in \mathbb{C}^{N \times 1}$, we have
	\begin{equation}\label{eq18}
		\begin{aligned}
		{\bf z} &={\bf A} \odot {\bf B} {\bf w} \\
		&=[{\bf b}_1 diag({{\bf a}_1}),\cdots,{\bf b}_N diag({{\bf a}_M})]^T\bf w \\
		&= [{\bf a}_1 diag({{\bf b}_1}),\cdots,{\bf a}_N diag({{\bf b}_M})]^T \bf w,
		\end{aligned}
	\end{equation}
\end{lemma}

\emph{Proof}: According to the Hadamard product, we can rewrite ${\bf A} \odot {\bf B}=[{\bf a}_1\odot{{\bf b}_1},\cdots,{{\bf a}_M\odot{\bf b}_M}]^T$. Thence, ${ z}_i $ is derived by $ ({\bf a}_i\odot{{\bf b}_i}){\bf w}$. Consider the Hadamard product of two vectors, it can be rewritten as a vector times a diagonal matrix as ${\bf a}_i\odot{{\bf b}_i}={\bf a}_i diag({{\bf b}_i})$ or ${\bf b}_i diag({{\bf a}_i})$. Q.E.D.

Along with the Lemma 2, ${{\boldsymbol{\Phi }}_k}({{\boldsymbol{\theta }}_z}) \odot {\bf{A}}({{\boldsymbol{\theta }}_z})= [{{{\bf{a}}_1({{\boldsymbol{\theta }}_z})}}diag({{{\boldsymbol{\phi }}_{1,k}}({{\boldsymbol{\theta }}_z})} ,\cdots,{{{\bf{a}}_M({{\boldsymbol{\theta }}_z})}} diag({{{\boldsymbol{\phi }}_{M,k}}({{\boldsymbol{\theta }}_z})}]^T$. We choose the same pattern for all DMA in one time slot as  ${{{\boldsymbol{\phi }}_{i,k}}({{\boldsymbol{\theta }}_z})}={{{\boldsymbol{\phi }}_{j,k}}({{\boldsymbol{\theta }}_z})}  \forall, i,j$, which can be extracted from the matrix. By substituting (\ref{eq13}) into (\ref{eq17}), the received signal is further transformed into
\begin{equation}\label{eq19}
{{\bf{Y}}_{k,t}} = \sum\limits_{z = a,j} {{\bf{A}}({{\boldsymbol{\theta }}_z})diag({\boldsymbol{\omega }}_{0,k}^H{{\bf{T}}}{\boldsymbol{\delta }}({{\boldsymbol{\theta }}_z})){{\bf{g}}_{zb}}{x_{z,t}} + {{\bf{n }}_k}},
\end{equation}
where ${\boldsymbol{\omega }}_{0,k}$ is the phase shifts of the DMA in $k$-th sampling. For $K_r$ observations, we can obtain a higher spatial resolution by reconstructing the signal space with different mapping (i.e., the radiation pattern).  As long as the sampling of different patterns is sufficient for the angular space, all the information on the angular spectrum can be recovered. Specifically, if $K_r=N$ and $N$-dimension Hadamard matrix exists, we can use it to construct a simple orthogonal matrix. For example, we set ${{\boldsymbol{\omega }}_{0,k}} = {{\bf{H}}_N}(k)$  where ${{\bf{H}}_N}(k)$ is the $k$-th column of the Hadamard matrix, which can be easily implemented by DMA with $1$bit control.

After obtaining $K_r $ observations, we use ${{\boldsymbol{\omega }}_{i}^H} = {{\bf{H}}_N^H}(i)$ for weighted combination of the signals and finally generate the $K_r$ virtual antennas, and the $i$-th set of signals ${{\bf{ Y}}_{i,t}^{'}} $ is
	\begin{equation}\label{eq19a}
	\begin{aligned}
& \sum\limits_{k = 1}^{K_r} {{\boldsymbol{\omega }}_{i}^{}(k)\sum\limits_{z = a,j} {{\bf{A}}({{\boldsymbol{\theta }}_z})diag({\boldsymbol{\omega }}_{0,k}^H{{\bf{T}}_0}{\boldsymbol{\delta }}({{\boldsymbol{\theta }}_z})){{\bf{g}}_{zb}}{x_{z,t}}} }  + {{\boldsymbol{\varepsilon }}_k}\\
&= \sum\limits_{z = a,j} {{\bf{A}}({{\boldsymbol{\theta }}_z})diag(\sum\limits_{k = 1}^{K_r} {{\boldsymbol{\omega }}_{i}^{}(k){\boldsymbol{\omega }}_{0,k}^H{{\bf{T}}_0}{\boldsymbol{\delta }}({{\boldsymbol{\theta }}_z}))} {{\bf{g}}_{zb}}{x_{z,t}}}  + {{\boldsymbol{\varepsilon }}_k}.
	\end{aligned}
\end{equation}
where ${{\boldsymbol{\varepsilon }}_k}$ is the mixed noise. According to the properties of Hadamard matrix, there is $\sum\limits_{k = 1}^{{K_r}} {{\boldsymbol{\omega }}_i^{}(k){\boldsymbol{\omega }}_k^H{{\bf{T}}_0}{\boldsymbol{\delta }}({{\boldsymbol{\theta }}_z})}  = {\boldsymbol{\omega }}_i^{}{\bf{H}}{{\bf{T}}_0}{\boldsymbol{\delta }}({{\boldsymbol{\theta }}_z}) = {K_r}{{\bf{p}}_i}{{\bf{T}}_0}{\boldsymbol{\delta }}({{\boldsymbol{\theta }}_z}) = {K_r}{t_{0,i}}{e^{ - jk_0 {r_i}\sin {{\boldsymbol{\theta }}_z} }}$, where all elements of the selection vector ${\bf p}_i$ are $0$ except that the $i$-th element is $1$. Thence, we finally get $K_r*M$ virtual antenna, and the signal of the $i$-th set of virtual antennas is
	\begin{equation}\label{eq19b}
	\begin{aligned}
&{{\bf{Y}}_{i,t}^{'}} = \sum\limits_{z = a,j} {{\bf{A}}({{\boldsymbol{\theta }}_z}){K_r}{t_{0,i}}diag({e^{ - jk_0 {r_i}\sin {{\boldsymbol{\theta }}_z} }}){{\bf{g}}_{zb}}{x_{z,t}}}  + {{\boldsymbol{\varepsilon }}_k}.
	\end{aligned}
\end{equation}

The resolution of the system is improved by combining multiple sampling results to generate multiple equivalent virtual antennas. Next, we give an estimate of the DoA by means of the ANM method. The atomic set is defined as the heterogeneous antenna manifold for all directions
\begin{equation}\label{eq20}
{{\cal A}} = \left\{ {{\bf{A}}^{'}(\theta ) \in {\mathbb{C}^{K_rN \times 1}}:\theta  \in \left[ { - \frac{\pi }{2},\frac{\pi }{2}} \right]} \right\},
\end{equation}
where ${\bf{A}}^{'}(\theta )={\bf{A}}({{{\theta }}}) \otimes  {\boldsymbol{\delta }}({{\theta }})$. Therefore, the atomic norm for these signals are
\begin{equation}\label{eq21}
	{\left\| {{{\bf S}_{z}}} \right\|_{{\cal A}}} = \mathop {\inf }\limits_{{\theta _l},{{{h}}_l}} \left\{ {{{\cal L}}:{{{\bf S}}_{z}} = \sum\limits_{l = 1}^{{\cal L}} {{\bf{A}}({\theta _l}){{\rm{h}}_l},{\theta _l} \in \left[ { - \frac{\pi }{2},\frac{\pi }{2}} \right]} } \right\},
\end{equation}
where ${{\rm{h}}_l}$ should be $K_rt_{0,i}{{\rm{g}}_{zb,l}}{x_{z,t}}$. The minimum number of ${\ell }_0$ norm means the minimum number of atoms that make up the whole signal space. The minimum atomic norm problem can be transformed into the following SDP problem\cite{Chi2015}
\begin{equation}\label{eq23}
	\begin{aligned}
		\begin{array}{l}
			\mathop {\min }\limits_{{\bf{Z}},{\bf{u}}} Tr({\bf{Z}}) + Tr({\bf{T}}({\bf{u}}))\\
			s.t. \left[ {\begin{array}{*{20}{c}}
					{\bf{Z}}&{{\bf{S }}_{z}^H},\\
					{{{\bf{S}}_{z}}}&{{\bf{T}}({\bf{u}})}
			\end{array}} \right] \succeq 0,
		\end{array}
	\end{aligned}
\end{equation}
which can be solved by the convex optimization methods such as the CVX toolbox. ${{\bf{T}}({\bf{u}})}$ is a Toplitz matrix which admits the following Vandermonde decomposition with ${\bf{T}}({\bf{u}}) = {\bf{A}}({\bf{\theta }}){\bf{\tilde{H}}}{{\bf{A}}^H}({\bf{\theta }})$. The DoA of all signals can be calculated by traditional DoA estimation algorithms, such as Multiple Signal Classification (MUSIC) scheme, which is not repeated here. Previously, we did not distinguish between legitimate signals and jamming signals. Thus the estimated angle contains both jamming's and signal's multipath. Generally, we can distinguish the jamming from the legitimate signal by its own characteristics, such as the difference in the pilot sequence or signal power, etc.

In the CSI estimation process, we denote the estimated legitimate signals' DoA as ${\tilde{\boldsymbol\theta}}_a$ and jamming signals' DoA as ${\tilde{\boldsymbol \theta}}_j$. 
Then bring them into the (\ref{eq19}) and we can obtain $K_r\times T$ observations of the CSI of the multipath channel. When the pilot sequences are known, the CSI of each path is the only unknown parameter, and it can be solved  by the least squares method based on the LS algorithm. It is worth noting that the CSI estimation accuracy is not only related to the SNR but also the DoA estimation results. The DoA can further enhance the estimation accuracy for CSI by forming the corresponding pattern. If the pilot sequences are unknown, the CSI can also be estimated by blind estimation methods\cite{InsungKang1999}.

\vspace{-0.3cm}
\subsection{Robust Anti-jamming scheme for SINR Maximization}

After obtaining the estimations, each heterogeneous antenna is designed for superimposing each signal multipath in phase and mutually weakening the jamming signal multipaths. According to the model of the RHA, the weights of the antennas and the phase shifts of the DMA elements are jointly designed to maximize the lower bound of receiver's SINR under imperfect DoA and CSI.

In the estimation process, the estimation results of DoA and CSI always contain estimation errors. According to the SNR of the signal, there exists a bounded angle uncertainty for each estimated angle, noted as ${{\boldsymbol{\bar \theta }}_z} = {{\boldsymbol{\tilde \theta }}_z} + \Delta {{\boldsymbol{\theta }}_z}$ where ${{\boldsymbol{ \bar \theta }}_z}$ is the actual DoA, ${{\boldsymbol{\tilde \theta }}_z}$ is the estimated DoA, and $\Delta {{\boldsymbol{\theta }}_z}$ is the DoA estimation error with $\Vert\Delta {{\boldsymbol{\theta }}_z}\Vert \le {\rho _{{{\theta }},z}}$. Similarly, the CSI uncertainty is modeled as ${{\bf{\bar g}}_{zb}} = {{\bf{\tilde g}}_{zb}} + \Delta {{\bf{g}}_{zb}}$, where ${{\bf{\bar g}}_{zb}}$ is the actual CSI, $ {{\bf{\tilde g}}_{zb}}$ is the estimated CSI and $\Delta {{\bf{g}}_{zb}}$ is the estimation error of the CSI with $\Vert\Delta {{\bf{g}}_{zb}}\Vert \le {\rho _{g,z}}$.  Although the DoA angular estimation has a cumulative error effect on the CSI error's variance, the error distribution of the DoA is independent of the distribution of the CSI error.

Thence, we aim to maximize the lower bound of the received SINR by designing the phase shifts of the DMA elements and antenna weights. The problem for maximization of the SINR under uncertainty is formulated as
\begin{subequations}\label{eq35}
	\begin{gather}
		\begin{aligned}
&{\cal P}_1: 	\mathop {\max }\limits_{{\bf{w}},{\boldsymbol{\omega }}} \mathop {\min }\limits_{\Delta {{\bf{g}}_{zb}},\Delta {{{\theta }}_z}} \frac{{{P_a}|{{\bf{w}}^H}{\boldsymbol{\Phi }}({{{\boldsymbol{\bar \theta }}}_a}) \odot {\bf{A}}({{{\boldsymbol{\bar \theta }}}_a}){{{\bf{\bar g}}}_{ab}}{|^2}}}{{{P_j}|{{\bf{w}}^H}{\boldsymbol{\Phi }}({{{\boldsymbol{\bar \theta }}}_j}) \odot {\bf{A}}({{{\boldsymbol{\bar \theta }}}_j}){{{\bf{\bar g}}}_{jb}}{|^2} + \sigma _n^2}},\\
&\qquad \quad\qquad s.t. \ |{{\rm w}_m}| = 1,m = 1 \cdots M,\\
&\qquad \qquad \quad\Vert\Delta {{\bf{g}}_{zb}}\Vert \le {\rho _{g,z}}, \Vert\Delta {{\boldsymbol{\theta }}_z}\Vert \le {\rho _{{{\theta }},z}},\\
&|{{\boldsymbol{\omega }}_m}^H{\bf T}{\boldsymbol{\delta }}({{{\boldsymbol{\bar \theta }}}_j}){{{\bf{\bar g}}}_{jb}}{|^2} \le \xi |{{\boldsymbol{\omega }}_m}^H{\bf T}{\boldsymbol{\delta }}({{{\boldsymbol{\bar \theta }}}_a}){{{\bf{\bar g}}}_{ab}}|,m = 1 \cdots M,\\
&{\omega _{m,n}} = \left\{ {{e^{j\psi }},\psi  \in \Psi } \right\},n = 1 \cdots N,m = 1 \cdots M.
		\end{aligned}
	\end{gather}
\end{subequations}
where the third inequality constrains every antenna's jamming-to-signal ratio in order to retain the legitimate signals, and $\xi$ is a system parameter related to the LSB and quantization bits of the ADC. It is evident that ${\cal P}_1$ is a non-convex problem due to the quadratic fractional form coupled with the Hadamard product, and is further complicated by the infinite constraints caused by the estimation errors.

To tackle the difficulty and complexity of dealing with the discrete phase shift, we propose the closest optimal scheme by approximating the continuous phase shift. From now on, the received power of the signal and jamming in the objective function can be transformed into 
	\begin{equation}\label{eq43}
		\begin{aligned}
			&|{{\bf{w}}^H}{\boldsymbol{\Phi }}({{\boldsymbol{\bar \theta }}_z}) \odot {\bf{A}}({{\boldsymbol{\bar \theta }}_z}){{\bf{\bar g}}_{ab}}|\\
			&\mathop = \limits^{(a)}  \vert \sum\limits_{m = 1}^M {{\boldsymbol{\omega }}_m^H}{{{\bf T}{\boldsymbol \delta}({{\boldsymbol{\bar \theta }}_z}) }} diag({{\bf{a}}_m^H}({{\boldsymbol{\bar \theta }}_z}))({{\bf{\tilde g}}_{zb}}
			\! +\!  \Delta {{\bf{g}}_{zb}}) \vert,
		\end{aligned}
	\end{equation}
where (a) is derived from the second equation of Lemma \ref{lem2} and merging the weighted values of the array elements into the DMA phase shifts. 
\\

Next, we transform the estimation errors into more manageable forms. For the pattern error, it can be simplified as
\begin{equation}\label{eq36}
	\begin{aligned}
&{\boldsymbol{\delta }}({{{\boldsymbol{\tilde \theta }}}_z} + \Delta {{\boldsymbol{\theta }}_z})\\
&= {[{e^{ - jk_0 {r_1} \sin ({{{\boldsymbol{\tilde \theta }}}_z} + \Delta {{{\boldsymbol{\tilde \theta }}}_z})}}, \cdots ,{e^{ - jk_0 {r_N} \sin ({{{\boldsymbol{\tilde \theta }}}_z} + \Delta {{{\boldsymbol{\tilde \theta }}}_z})}}]^T}\\
&\mathop  \approx \limits^{(a)} {\boldsymbol\delta}({{{\boldsymbol{\tilde \theta }}}_z}) \odot {[{e^{ - jk_0 {r_1} \Delta {{{\boldsymbol{\tilde \theta }}}_z}}}, \cdots ,{e^{ - jk_0 {r_N} \Delta {{{\boldsymbol{\tilde \theta }}}_z}}}]^T},
	\end{aligned}
\end{equation}
where (a) is approximated with $\sin (\theta+\Delta \theta)=\sin (\theta)\cos (\Delta \theta) + \cos (\theta)\sin (\Delta \theta) \approx \sin (\theta) + \Delta \theta $ when $\Delta \theta$ is small.
	Similarly, we can rewrite the array steering errors as
	\begin{equation}\label{eq37}
		\begin{aligned}
		&{{\bf{a}}_m}({{{\boldsymbol{\tilde \theta }}}_z} + \Delta {{{\boldsymbol{\tilde \theta }}}_z}) \approx {{\bf{a}}_m}({{{\boldsymbol{\tilde \theta }}}_z}) \odot {[{e^{ - jk_0 {d_m} \Delta {{{\boldsymbol{\tilde \theta }}}_z}}}]^T}.
		\end{aligned}
	\end{equation}

By unifying the errors, the coupling part in (\ref{eq43}) can be transformed into
\begin{equation}\label{eq38}
	\begin{aligned}
&	{\boldsymbol{\delta }}({{{\boldsymbol{\bar \theta }}}_z})diag({{\boldsymbol{a}}_m}^H({{{\boldsymbol{\bar \theta }}}_z}))\\
&= ({\boldsymbol \Xi}_m (\Delta {{\boldsymbol{\theta }}_z}) \odot {\boldsymbol{\delta }}({{{\boldsymbol{\tilde \theta }}}_z}))diag({{\bf{a}}_m}^H({{{\boldsymbol{\tilde \theta }}}_z})),
	\end{aligned}
\end{equation}
where ${\boldsymbol \Xi}_m (\Delta {{\boldsymbol{\theta }}_z}) = {[{e^{ - jk_0 ({r_1} + {d_m}) \Delta {{{\boldsymbol{\tilde \theta }}}_z}}}, \cdots ,{e^{ - j2k_0 ({r_N} + {d_m}) \Delta {{{\boldsymbol{\tilde \theta }}}_z}}}]^T}$. Apparently, the multiplicative error matrix caused by the DoA estimation error is difficult to optimize. Here we approximate the error with Taylor expansions as
\begin{equation}\label{eq39}
	\begin{aligned}
	{e^{ - jk_0 ({r_n} + {d_m})\Delta {{\boldsymbol{\theta }}_z} }}= 1 + j k_0 ({r_n} + {d_m}) \Delta {{\boldsymbol{\theta }}_z} + {{\boldsymbol{\varepsilon }}_{\boldsymbol{\delta }}},
	\end{aligned}
\end{equation}
where ${{\boldsymbol{\varepsilon }}_{\boldsymbol{\delta }}} = \sum\limits_{n = 2}^\infty  {\frac{{{{( - jk_0 ({r_n} + {d_m}) \Delta {{\boldsymbol{\theta }}_z})}^n}}}{{n!}}} $.  The subsequent high order quantities gradually converge to $0$ and  can be omitted. The combined error matrix is transformed into
\begin{equation}\label{eq40}
	\begin{aligned}
&	{\boldsymbol \Xi} (\Delta {{\bf{\theta }}_z})\\
&= {{\bf{E}}_{N \times L}} + [j2\pi ({r_1} + {d_m})/\lambda \Delta {{\boldsymbol{\theta }}_z},j2\pi ({r_N} + {d_m})/\lambda \Delta {{\boldsymbol{\theta }}_z}]\\
&= {{\bf{E}}_{N \times L}} + {[j2\pi ({r_1} + {d_m})/\lambda ,j2\pi ({r_N} + {d_m})/\lambda ]^T}\Delta {{\boldsymbol{\theta }}_z},
	\end{aligned}
\end{equation}
where ${{\bf{E}}_{N \times L}}$ is the full $1$ matrix.  By substituting (\ref{eq40}) into (\ref{eq38}), we have
\begin{equation}\label{eq41}
	\begin{aligned}
	{\boldsymbol \Xi}_m (\Delta {{\boldsymbol{\theta }}_z}) \odot {\boldsymbol{\delta }}({{{\boldsymbol{\tilde \theta }}}_z})= {\boldsymbol{\delta }}({{{\boldsymbol{\tilde \theta }}}_z}) + \Delta {\boldsymbol{\delta }}_m({{{\boldsymbol{\tilde \theta }}}_z}).
	\end{aligned}
\end{equation}

Note that the length of each heterogeneous antenna will not be greater than the spacing between the arrays, so ${r_1} <  \cdots  < {r_N} \le {d_1}$. By relaxing the estimation error of the DMA elements, the equivalent pattern error is defined as 
\begin{equation}\label{eq42}
	\begin{aligned}
		\Delta {\boldsymbol{\delta }}_m({{{\boldsymbol{\tilde \theta }}}_z}) =
		{\boldsymbol{\delta }}({{{\boldsymbol{\tilde \theta }}}_z}){{\bf{R}}_{z,m}}diag(\Delta {{\boldsymbol{\theta }}_z}),
	\end{aligned}
\end{equation}
with ${{\bf{R}}_{z,m}} = j{k_0}{d_{m + 1}}{\bf I}_{L_z\times L_z}$.  For convenience, we straighten the phase shift vectors into one vector, i.e. ${{\bf{v}}} = [{{\boldsymbol{\omega }}_1}^H, \cdots ,{{\boldsymbol{\omega }}_M}^H]^H \in {\mathbb{C}^{NM \times 1}}$. Similarly, we use the characteristic of the block matrix and rewrite the matrices in (\ref{eq43}) as ${\bf \widehat{T}}  = blockdiag(\underbrace {{\bf{T}}, \cdots ,{\bf{T}}}_M) \in {\mathbb{C}^{NM \times NM}}$,   ${\bf{\widehat{A}}}({{\boldsymbol{\theta }}_z}) = diag([{{\bf{\tilde a}}_1}^H({{\boldsymbol{\theta }}_z}), \cdots ,{{\bf{\tilde a}}_M}^H({{\boldsymbol{\theta }}_z})]) \in {\mathbb{C}^{M{L_z} \times M{L_z}}}$ and ${{\bf{G}}_{zb}} = {[\underbrace {{{{\bf{\tilde g}}}_{zb}}^H, \cdots ,{{{\bf{\tilde g}}}_{zb}}^H}_M]^H} \in {\mathbb{C}^{M{L_z} \times 1}}$. The RHA steering matrix is ${\bf{D}}({{\boldsymbol{\theta }}_z}) = blockdiag(\underbrace {{\boldsymbol{\delta }}({{\boldsymbol{\theta }}_z}), \cdots ,{\boldsymbol{\delta }}({{\boldsymbol{\theta }}_z})}_M) \in {\mathbb{C}^{NM \times M{L_z}}}$ and steering error matrix$\Delta {\bf{D}}({{\bf{\theta }}_z}) \in {\mathbb{C}^{NM \times M{L_z}}}$ is defined by
		\begin{equation}\label{eq44}
		\begin{aligned}
&\Delta {\bf{D}}({{\bf{\theta }}_z})= diag(\underbrace {\Delta {\boldsymbol{\delta }}_1({{\boldsymbol{\theta }}_z}), \cdots ,\Delta {\boldsymbol{\delta }}_M({{\boldsymbol{\theta }}_z})}_M) ={\bf{D}}({{\boldsymbol{\theta }}_z})  {\bf \widehat R}_z{\Delta {{\boldsymbol{\hat \theta }}_z}}
		\end{aligned}
	\end{equation}
where ${\bf \widehat R}_z= blockdiag({{\bf{R}}_{z,1}}, \cdots {{\bf{R}}_{z,M}})$ and ${\Delta {{\boldsymbol{\hat \theta }}_z}}={{\bf I}_{M \times 1}} \otimes diag(\Delta {{\boldsymbol{\theta }}_z})$. And the CSI error matrix is extended as  $\Delta {{\bf{G}}_{zb}} = {[\Delta \underbrace {{{{\bf{\tilde g}}}_{zb}}^H, \cdots ,\Delta {{{\bf{\tilde g}}}_{zb}}^H}_M]^H} \in {\mathbb{C}^{M{L_z} \times 1}}$.

By introducing two variables with in (\ref{eq46}) and (\ref{eq47}), ${\cal P}_{1}$ is transformed into (\ref{eq48}).
\begin{figure*}
	\vspace{-0.4cm}
	\begin{equation}\label{eq46}
		\begin{aligned}
			\gamma  = \mathop {\min }\limits_{\Delta {{\boldsymbol{\theta }}_z},\Delta {{\bf{G}}_{zb}}} \frac{{{P_a}|{{\bf{v}}^H}{\bf \widehat{T}} ({\bf{D}}({{\boldsymbol{\theta }}_a}) + \Delta {\bf{D}}({{\boldsymbol{\theta }}_a})){\bf{\widehat A}}({{\boldsymbol{\theta }}_a})({{\bf{G}}_{ab}} + \Delta {{\bf{G}}_{ab}}){|^2}}}{{{P_j}|{{\bf{v}}^H}{\bf \widehat{T}} ({\bf{D}}({{\boldsymbol{\theta }}_j}) + \Delta {\bf{D}}({{\boldsymbol{\theta }}_j})){\bf{\widehat A}}({{\boldsymbol{\theta }}_j})({{\bf{G}}_{jb}} + \Delta {{\bf{G}}_{jb}})| + \sigma _n^2}}
		\end{aligned}
	\end{equation}
\end{figure*}

\begin{figure*}
	\vspace{-0.4cm}
	\begin{equation}\label{eq47}
		\begin{aligned}
			b = \mathop {\max }\limits_{\Delta {{\boldsymbol{\theta }}_j},\Delta {{\bf{G}}_{jb}}} {P_j}|{{\bf{v}}^H}{\bf \widehat{T}} ({\bf{D}}({{\boldsymbol{\theta }}_j}) + \Delta {\bf{D}}({{\boldsymbol{\theta }}_j})){\bf{\widehat A}}({{\boldsymbol{\theta }}_j})({{\bf{G}}_{jb}} + \Delta {{\bf{G}}_{jb}})|	
		\end{aligned}
	\end{equation}
\end{figure*}

	\begin{figure*}
		\vspace{-0.7cm}
			\begin{subequations}\label{eq48}
					\begin{gather}
							\begin{aligned}
			&\qquad\qquad\qquad\qquad\qquad\mathop {\max }\limits_{{\bf{v}}}  \gamma \\
s.t.&\mathop {\min }\limits_{\Delta {{\boldsymbol{\theta }}_a},\Delta {{\bf{G}}_{ab}}} {P_a}|{{\bf{v}}^H}{\bf \widehat{T}} ({\bf{D}}({{\boldsymbol{\theta }}_a}) \!+\! \Delta {\bf{D}}({{\boldsymbol{\theta }}_a})){\bf{\widehat A}}({{\boldsymbol{\theta }}_a})({{\bf{G}}_{ab}}\! + \!\Delta {{\bf{G}}_{ab}}){|^2}  \ge \gamma b\\
&\mathop {\max }\limits_{\Delta {{\boldsymbol{\theta }}_j},\Delta {{\bf{G}}_{jb}}} {P_j}|{{\bf{v}}^H}{\bf \widehat{T}} ({\bf{D}}({{\boldsymbol{\theta }}_j}) + \Delta {\bf{D}}({{\boldsymbol{\theta }}_j})){\bf{\widehat A}}({{\boldsymbol{\theta }}_j})({{\bf{G}}_{jb}}\! +\! \Delta {{\bf{G}}_{jb}}){|^2}  \le b\!-\! \sigma _n^2,\\
&\mathop {\max }\limits_{\Delta {{\boldsymbol{\theta }}_z},\Delta {{\bf{G}}_{zb}}} |{{\bf{v}}^H}{{\bf{P}}_m}{\bf \widehat{T}} ({\bf{D}}({{\bf{\theta }}_j}) + \Delta {\bf{D}}({{\boldsymbol{\theta }}_j})){\bf{\widehat A}}({{\boldsymbol{\theta }}_j})({{\bf{G}}_{jb}} + \Delta {{\bf{G}}_{jb}})|\nonumber\\
& \le \mathop {\min }\limits_{\Delta {{\boldsymbol{\theta }}_z},\Delta {{\bf{G}}_{zb}}} \xi |{{\bf{v}}^H}{{\bf{P}}_m}{\bf \widehat{T}} ({\bf{D}}({{\boldsymbol{\theta }}_a}) + \Delta {\bf{D}}({{\boldsymbol{\theta }}_a})){\bf{\widehat A}}({{\boldsymbol{\theta }}_a})({{\bf{G}}_{ab}} + \Delta {{\bf{G}}_{ab}})|,m = 1 \cdots M\\
& \quad\qquad \qquad\qquad\vert {v_n}\vert = 1,n = 1 \cdots NM,\\
& \quad \qquad\qquad\Vert \Delta {{\bf{G}}_{zb}}\Vert \le \sqrt M {\rho _{g,z}},\Vert\Delta {{\bf{\theta }}_z}\Vert \le {\rho _{{\bf{\delta }},z}}.
								\end{aligned}
						\end{gather}
				\end{subequations}
\hrulefill
		\end{figure*}

Then, we relax the angle estimation error in constraint (\ref{eq48}b). There is
	\begin{equation}\label{eq49}
	\begin{aligned}
	&\mathop {\min }\limits_{\Delta {\boldsymbol{\theta }}} |{{\bf{v}}^H}{\bf \widehat{T}} ({\bf{D}}({{\boldsymbol{\theta }}_a}) + \Delta {\bf{D}}({{\boldsymbol{\theta }}_a})){\bf{\widehat A}}({{\boldsymbol{\theta }}_a})({{\bf{G}}_{ab}} + \Delta {{\bf{G}}_{ab}}){|^2}\\
	&= \mathop {\min }\limits_{\Delta {\boldsymbol{\theta }}} |{{\bf{v}}^H}{\bf \widehat{T}} {\bf{D}}({{\boldsymbol{\theta }}_a}){\bf{\widehat A}}({{\boldsymbol{\theta }}_a})({{\bf{G}}_{ab}} + \Delta {{\bf{G}}_{ab}}) \\
	&\quad + {{\bf{v}}^H}{\bf \widehat{T}} {\bf D}({{\boldsymbol{\theta }}_a}){\bf{\widehat R}}_a\Delta {\boldsymbol{\hat \theta }}_a{\bf{\widehat A}}({{\boldsymbol{\theta }}_a})({{\bf{G}}_{ab}} + \Delta {{\bf{G}}_{ab}}){|^2}.
	\end{aligned}
\end{equation}
In particular, the order can be swapped with the subsequent diagonal matrix as
\begin{equation}\label{eq50}
	\begin{aligned}
&{{\bf{v}}^H}{\bf \widehat{T}}{\bf D}({{\boldsymbol{\theta }}_a}){\bf{\widehat R}}_a\Delta {\boldsymbol{\hat \theta }}_a{\bf{\widehat A}}({{\boldsymbol{\theta }}_a})({{\bf{G}}_{ab}} + \Delta {{\bf{G}}_{ab}})\\
&={{\bf{v}}^H}{\bf \widehat{T}} {\bf{D}}({{\boldsymbol{\theta }}_a}){\bf{\widehat A}}({{\boldsymbol{\theta }}_a}){\bf{\widehat R}}_a(diag({{\bf{G}}_{ab}})\Delta {\boldsymbol{\hat \theta }}+diag(\Delta{{\bf{G}}_{ab}})\Delta {\boldsymbol{\hat \theta }}_a)\\
&\mathop  \approx \limits^{(a)} {{\bf{v}}^H}{\bf \widehat{T}} {\bf{D}}({{\boldsymbol{\theta }}_a}){\bf{\widehat A}}({{\boldsymbol{\theta }}_a}){\bf{\widehat R}}_a diag({{\bf{G}}_{ab}})\Delta {\boldsymbol{\hat \theta }_a},
	\end{aligned}
\end{equation}
where (a) comes from the fact that the terms multiplied by the two error terms are relatively small and thus are ignored. By substituting (\ref{eq50}) into the second part,  (\ref{eq49}) is converted into
\begin{equation}\label{eq51}
	\begin{aligned}
	\mathop {\min }\limits_{\Delta {\boldsymbol{\theta }_a}}\vert{{\bf{v}}^H}{\bf \widehat{T}} {\bf{D}}({{\boldsymbol{\theta }}_a}){\bf{\widehat A}}({{\boldsymbol{\theta }}_a})({{\bf{G}}_{ab}} \!+\! \Delta {{\bf{G}}_{ab}}\!+\!{\bf{\widehat R}}_a diag({{\bf{G}}_{ab}})\Delta {\boldsymbol{\hat \theta }_a})\vert^2.
	\end{aligned}
\end{equation}

Similarly, the constraints in (\ref{eq48}b), (\ref{eq48}c) and (\ref{eq48}d) are transformed into
		\begin{subequations}\label{eq52}
	\begin{gather}
		\begin{aligned}
&		\mathop {\min }\limits_{\Delta {{\bf{G}}_{ab}},\Delta {\boldsymbol{ \theta }}_a} {P_a}|{{\bf{v}}^H}{\bf \widehat{T}} {\bf{D}}({{\boldsymbol{\theta }}_a}){\bf{\widehat A}}({{\boldsymbol{\theta }}_a})({{\bf{G}}_{ab}} \!+\! \Delta {{\bf{G}}_{ab}}\!+\!\Delta {\bf \tilde G}_{a}){|^2} \nonumber\\
&\qquad\qquad\qquad\qquad\qquad\qquad\qquad\qquad\ge \gamma b,\\
&\mathop {\max }\limits_{\Delta {{\bf{G}}_{jb}},\Delta {\boldsymbol{ \theta }}_j} {P_j}|{{\bf{v}}^H}{\bf \widehat{T}} {\bf{D}}({{\boldsymbol{\theta }}_j}){\bf{\widehat A}}({{\boldsymbol{\theta }}_j})({{\bf{G}}_{jb}} + \Delta {{\bf{G}}_{jb}}\!+\!\Delta {\bf \tilde G}_{j}){|^2}  \nonumber\\
&\qquad\qquad\qquad\qquad\qquad\qquad\qquad\qquad\le b- \sigma _n^2,\\
&\mathop {\max }\limits_{\Delta {{\bf{G}}_{jb}},\Delta {\boldsymbol{ \theta }}_j } |{{\bf{v}}^H}{{\bf{P}}_m}{\bf \widehat{T}} {\bf{D}}({{\boldsymbol{\theta }}_j}){\bf{\widehat A}}({{\boldsymbol{\theta }}_j})({{\bf{G}}_{jb}} + \Delta {{\bf{G}}_{jb}}\!+\!\Delta {\bf \tilde G}_{j})|\le\nonumber\\
& \mathop {\min }\limits_{\Delta {{\bf{G}}_{ab}},\Delta {\boldsymbol{ \theta }}_a} \xi |{{\bf{v}}^H}{{\bf{P}}_m}{\bf \widehat{T}} {\bf{D}}({{\boldsymbol{\theta }}_a}){\bf{\widehat A}}({{\boldsymbol{\theta }}_a})({{\bf{G}}_{ab}} + \Delta {{\bf{G}}_{ab}}\!+\!\Delta {\bf \tilde G}_{a})|,\nonumber\\
&\qquad\qquad\qquad\qquad\qquad\qquad\qquad\qquad m = 1 \cdots M.
		\end{aligned}
	\end{gather}
\end{subequations}
where $\Delta {\bf \tilde G}_{z}={\bf{\widehat R}}_z diag({{\bf{G}}_{zb}})\Delta {\boldsymbol{\hat \theta }}_z)$.
 Then we introduce a SDR variable with ${\bf V}={\bf vv}^H$, and the constraint (\ref{eq52}a) is equal to
	\begin{equation}\label{eq53}
		\begin{aligned}
			& \vert {{\bf{v}}^H}{\bf \widehat{T}} {\bf{D}}({{\boldsymbol{\theta }}_a}){\bf{\widehat A}}({{\boldsymbol{\theta }}_a})({{\bf{G}}_{ab}} + \Delta {{\bf{G}}_{ab}}\!+\!\Delta {\bf \tilde G}_{a}){\vert^2}\\
			&=\Delta {{\bf{G}}_{ab}}^H{{\bf{Q}}_{ab}}\Delta {{\bf{G}}_{ab}} + {	({{\bf{G}}_{ab}}\!+\!\Delta {\bf \tilde G}_{a})}^H{{\bf{Q}}_{ab}}\Delta {{\bf{G}}_{ab}} \\
			& + \Delta {{\bf{G}}_{ab}}^H{{\bf{Q}}_{ab}}{	({{\bf{G}}_{ab}}\!+\!\Delta {\bf \tilde G}_{a})} +c_{ab}	,
		\end{aligned}
	\end{equation}
where ${{\bf{Q}}_{ab}} = {\bf{\widehat A}}{({{\boldsymbol{\theta }}_a})^H}{\bf{D}}{({{\boldsymbol{\theta }}_a})^H}{{\bf \widehat{T}} ^H}{\bf{V}}{\bf \widehat{T}} {\bf{D}}({{\boldsymbol{\theta }}_a}){\bf{\widehat A}}({{\boldsymbol{\theta }}_a})$ and $c_{ab}={({{\bf{G}}_{ab}}\!+\!\Delta {\bf \tilde G}_{a})}^H{{\bf{Q}}_{ab}}{({{\bf{G}}_{ab}}\!+\!\Delta {\bf \tilde G}_{a})}$. Further, we introduce S-procedure for handling the norm-constraint problems.

\begin{lemma}\label{lem3}
	(S-procedure) There are a set of functions with ${g_k}(x) = {{\bf{x}}^H}{{\bf{Q}}_k}{\bf{x}} + {\bf{q}}_{}^H{\bf{x}} + {\bf{x}}_{}^H{\bf{q}} + c, k\in{1,2}$, for the Hermitian matrix ${{\bf{Q}}_k} \in \mathbb{C}{^{L \times L}}$, ${\bf{q}} \in {\mathbb{C}^L}$ and $c \in \mathbb{R}$. Then, the implication ${g_2}({\bf \hat x}) \le 0 \Rightarrow  {g_1}( {\bf \hat x} ) \le 0$ holds if and only if there exists $\alpha  \ge 0$ satisfying
\begin{equation}\label{eq54}
\alpha \left[ {\begin{array}{*{20}{c}}
		{{{\bf{Q}}_1}}&{{{\bf{q}}_1}}\\
		{{\bf{q}}_1^H}&{{c_1}}
\end{array}} \right] - \left[ \begin{array}{*{20}{c}}
{{\bf{Q}}_2}&{{\bf{q}}_2}\\
{{\bf{q}}_2^H}&{c_2}
\end{array}
\right] \succeq 0.
	\end{equation}
\end{lemma}

Based on  Lemma \ref{lem3}, by defining (\ref{eq53}) and  (\ref{eq48}f) as $g_1(\Delta {{\bf{G}}_{ab}})$ and $g_2(\Delta {{\bf{G}}_{ab}})$, respectively, we have
\begin{equation}\label{eq55}
	\begin{aligned}
&\left[ \begin{array}{*{20}{c}}
	{\alpha_1 {{\bf{I}}_{M \times {L_a}}} + {{\bf{Q}}_{ab}}} & {{{\bf{Q}}_{ab}}({{\bf{G}}_{ab}}\!+\!\Delta {\bf \tilde G}_{a})}\\
	{({{\bf{G}}_{ab}}\!+\!\Delta {\bf \tilde G}_{a})}^H {{\bf{Q}}_{ab}} & {\begin{array}{c}
			c_{ab} - \gamma b/({P_a}) - \alpha_1 {\rho ^2_{g,a}} \end{array}}
\end{array} \right]\succeq{{0}},\\
&\qquad\qquad\qquad\qquad \Vert\Delta {{\boldsymbol{\theta }}_a}\Vert \le {\rho _{{{\theta }},a}}.
	\end{aligned}
\end{equation}

To handle the angular error, we introduce Lemma \ref{lem4} as

\begin{lemma}\label{lem4}
	 For ${\bf D}\succeq 0$ and  ${\bf H}_j,j=1,\cdots,6$ satisfy the following inequality
	 \begin{equation}\label{eq56}
\left\{ \begin{array}{l}
	\left[ {\begin{array}{*{20}{c}}
			{{{\bf{H}}_1}}&{{{\bf{H}}_2} + {{\bf{H}}_3}{\bf{X}}}\\
			{{{({{\bf{H}}_2} + {{\bf{H}}_3}{\bf{X}})}^H}}&{\begin{array}{c}
					{{\bf{H}}_4} + {{\bf{X}}^H}{{\bf{H}}_5} \\
					+ {\bf{H}}_5^H{\bf{X}} + {{\bf{X}}^H}{{\bf{H}}_6}{\bf{X}}
			\end{array}}
	\end{array}} \right] \succeq 0,\\
	\forall {\bf{X}}:{\bf{I}} - {{\bf{X}}^H}{\bf{DX}} \succeq 0,
\end{array} \right.
	 \end{equation}
They are equivalent to the following inequality
\begin{equation}\label{eq57}
	 \left[\begin{array}{*{20}{c}}
	 	{{\bf H}_1}&{{\bf H}_2}&{{\bf H}_3}\\
	 	{{\bf H}_2^H}&{{\bf H}_4}&{{\bf H}_5^H}\\
	 	{{\bf H}_3^H}&{{\bf H}_5}&{{\bf H}_6}
	 \end{array}\right] - {\beta}\left[\begin{array}{*{20}{c}}
	 	{\bf 0}&{\bf 0}&{\bf 0}\\
	 	{\bf 0}&{\bf I}&{\bf 0}\\
	 	{\bf 0}&{\bf 0}&{\bf D}
	 	\end{array}\right]\succeq 0.
\end{equation}
where $\beta$ is an auxiliary variable, $\bf 0$ and $\bf I$ are zero matrix and identity matrix with specific dimensions.
\end{lemma}

Using Lemma \ref{lem4}, the constraint (\ref{eq55}) is further transformed into (\ref{eq58}). Similarly, the constraint (\ref{eq52}b) is identical to (\ref{eq59}).

\begin{figure*}
		\vspace{-0.7cm}
	\begin{equation}\label{eq58}
		\begin{aligned}
\left[ {\begin{array}{*{20}{c}}
		{\alpha_1 {\bf{I}} + {{\bf{Q}}_{ab}}}&{{{\bf{Q}}_{ab}}{{\bf{G}}_{ab}}}&{{{\bf{Q}}_{ab}}{\bf{\hat R}}diag({{\bf{G}}_{ab}})}\\
		{{\bf{G}}_{ab}^H {{\bf{Q}}_{ab}}}&{{\bf{G}}_{ab}^H{{\bf{Q}}_{ab}}{{\bf{G}}_{ab}} - \gamma b/({P_a}) - \alpha_1 {\rho _{g,a}}^2 - \beta_1 {\rho _{\theta ,a}}^2}&{{\bf{G}}_{ab}^H{{\bf{Q}}_{ab}}{\bf{\hat R}}diag({{\bf{G}}_{ab}})}\\
		{diag({\bf{G}}_{ab}^H){\bf{\hat RQ}}_{ab}^{}}&{diag({\bf{G}}_{ab}^H){\bf{\hat R}}{{\bf{Q}}_{ab}}{{\bf{G}}_{ab}}}&{diag({\bf{G}}_{ab}^H){\bf{\hat R}}{{\bf{Q}}_{ab}}{\bf{\hat R}}diag({{\bf{G}}_{ab}}) - \beta_1 {\bf{I}}}
\end{array}} \right]\succeq 0.
		\end{aligned}
	\end{equation}
				
\end{figure*}

 \begin{figure*}
 	\vspace{-0.7cm}
 	\begin{equation}\label{eq59}
 		\begin{aligned}
 			\left[ {\begin{array}{*{20}{c}}
 					{{\alpha _2}{\bf{I}} - {{\bf{Q}}_{jb}}}&{ - {{\bf{Q}}_{jb}}{{\bf{G}}_{jb}}}&{ - {{\bf{Q}}_{jb}}{\bf{\hat R}}diag({{\bf{G}}_{jb}})}\\
 					{ - {\bf{G}}_{jb}^H {{\bf{Q}}_{jb}}}&{b/{P_j} - {\bf{G}}_{jb}^H{{\bf{Q}}_{ab}}{{\bf{G}}_{jb}} - {\alpha _2}{\rho _{g,j}}^2 - {\beta _2}{\rho _{\theta ,j}}^2}&{ - {\bf{G}}_{jb}^H{{\bf{Q}}_{jb}}{\bf{\hat R}}diag({{\bf{G}}_{jb}})}\\
 					{ - diag({\bf{G}}_{jb}^H){\bf{\hat RQ}}_{jb}^{}}&{ - diag({\bf{G}}_{jb}^H){\bf{\hat RQ}}_{jb}^{}{{\bf{G}}_{jb}}}&{ - diag({\bf{G}}_{jb}^H){\bf{\hat R}}{{\bf{Q}}_{jb}}{\bf{\hat R}}diag({{\bf{G}}_{jb}}) - {\beta _2}{\bf{I}}}
 			\end{array}} \right]\succeq 0.
 		\end{aligned}
 	\end{equation}
 \end{figure*}

For the constraint (\ref{eq52}c), the effect caused by array errors on one heterogeneous antenna can be normalized. Thus, the total equivalent error is defined as $\Delta {{\bf{G}}_{t,z}} = diag({{\bf{G}}_{zb}})\Delta {{\bf{\hat \theta }}_z} + \Delta {{\bf{G}}_{zb}}$ with $\left\| {\Delta {{\bf{G}}_{t,z}}} \right\| \le {\rho _{\theta ,z}}{\sigma _z} + {\rho _{g,z}}$ by bring the expectation of the norm of the path gain.  According to the Lemma \ref{lem3}, the constraint is transformed into
		\begin{equation}\label{eq60}
			\begin{aligned}
\left[ {\begin{array}{*{20}{c}}
		{{\eta _{1,m}}{\bf{I}} + {\bf{Q}}_{ab,m} }&{{\bf{Q}}_{ab,m} ({{\bf{G}}_{ab}})}\\
		{{{({{\bf{G}}_{ab}})}^H}{\bf{Q}}_{ab,m} }&{ {e_{m}} - {\eta _{1,m}}{{({\rho _{\theta ,a}}{\sigma _z} + {\rho _{g,a}})}^2}}
\end{array}} \right] \succeq {{0}},
			\end{aligned}
		\end{equation}			
where ${{\bf{Q}}_{zb,m}} = {\bf{\widehat A}}{({{\boldsymbol{\theta }}_z})^H}{\bf{D}}{({{\boldsymbol{\theta }}_z})^H}{{\bf P}_m^H}{{\bf \widehat{T}} ^H}{\bf{V}}{\bf \widehat{T}}{{\bf P}_m} {\bf{D}}({{\boldsymbol{\theta }}_z}){\bf{\widehat A}}({{\boldsymbol{\theta }}_z})$ and 	${e_{m}} = ({({{\bf{G}}_{jb}} + \Delta {{\bf{G}}_{t,j}})^H}{\bf{Q}}_{jb,m}({{\bf{G}}_{jb}} + \Delta {{\bf{G}}_{t,j}}))-\xi {{\bf{G}}_{ab}}^H{\bf{Q}}_{ab,m}{{\bf{G}}_{ab}}^H$. The LMI inequality is further rewritten as (\ref{eq61}) by Lemma \ref{lem4}, where ${C_m}  = \xi {{\bf{G}}_{ab}}^H{\bf{Q}}_{ab,m}{{\bf{G}}_{ab}} - {{\bf{G}}_{jb}}^H{\bf{Q}}_{jb,m}{{\bf{G}}_{jb}} - {\eta _{1,m}}{({\rho _{\theta ,a}}{\sigma _a} + {\rho _{g,a}})^2}$.

 \begin{figure*}
 	\vspace{-0.5cm}
	\begin{equation}\label{eq61}
		\begin{aligned}
\left[ {\begin{array}{*{20}{c}}
		{{\eta _{1,m}}{\bf{I}} + \xi {\bf{Q}}_{ab,m}}&{\xi {\bf{Q}}_{ab,m}{{\bf{G}}_{ab}}}&{\bf{0}}\\
		{\xi {\bf{G}}_{ab}^H {\bf{Q}}_{ab,m}}&{{C_m}  - {\eta _{2,m}}}&{ - {\bf{G}}_{jb}^H{\bf{Q}}_{jb,m}}\\
		{\bf{0}}&{ - {\bf{Q}}_{jb,m} {{\bf{G}}_{jb}}}&{\frac{{{\eta _{2,m}}}}{{{{({\rho _{\theta ,j}}{\sigma _j} + {\rho _{g,j}})}^2}}}{\bf{I}} - {\bf{Q}}_{jb,m}}
\end{array}} \right]\succeq 0, \forall m\in\{1,\cdots M \}.
		\end{aligned}
	\end{equation}
\hrulefill
\end{figure*}

Using the SDR method for the constraint (\ref{eq48}e), the optimal problem finally becomes
\begin{equation}\label{eq62}
	\begin{aligned}
    & \qquad\qquad {\cal P}_{1.2}: \quad \mathop  {\max }\limits_{{\bf{V}},{\boldsymbol{\alpha,\beta, \eta} } ,b}  \gamma ,\\
        & \vert{\bf{V}}_{n,n}\vert=1, n=1, \cdots, NM,  {\bf{V}} \succeq 0, rank({\bf{V}})=1,\\
    &\quad  \alpha_z ,\beta_z, \eta_{z,m} \ge 0, \forall z\in\{1,2 \}, \forall m\in\{1,\cdots M \},\\
    &\qquad\qquad\qquad (\ref{eq58}), (\ref{eq59}), (\ref{eq61}).
	\end{aligned}
\end{equation}		

Since the problem is non-convex caused by the  rank-$1$ constraint and coupling of target $\gamma $ and variable $b$, we propose an alternating optimization method to handle it. Without the rank-$1$ constraint, the problem ${\cal P}_{1.2}$ is an SDP problem that can be solved by the interior-point method with fixed $\gamma$. And the optimal SINR and the corresponding $V$ can be alternatively updated by a bisection search. When the obtained $\bf V$ is rank-$1$, the optimal continuous phase shifts can be  given directly by the eigenvalue decomposition. Otherwise, a Gaussian randomization method \cite{Wang2021} with sufficient sampling can be used to generate a set of rank-$1$ solutions that approximate the optimal value.

After that, we can map the continuous phase shifts to the discrete ones by the minimum norm distance. Since the discrete and continuous phase shifts are modulo 1, the distance between them can be described as the degree of offset in complex coordinates. Therefore, the optimal discrete phase shift for one element can be given by ${{\bar \omega }_{m,n}^{opt}} = \arg \mathop {\max }\limits_{{\omega _b}} |\omega _{m,n}^{opt} + {\omega _b}|$ for ${\omega _b} = \left\{ {{e^{j\psi }},\psi  \in \Psi } \right\}$. However, simple mapping can not give the best results. In fact, the discrete phase shifts consist of both element’s phase shift and weighted value on the antenna. The optimization problem is
\begin{equation}\label{eq63}
	\begin{aligned}
	&\mathop {\max }\limits_{{w_m},{{\bar \omega }_{m,n}}} {\bar E}=\sum\limits_{n = 1}^N {|\omega _{m,n}^{opt} + {w_m}{{\bar \omega }_{m,n}}|}, \\
&s.t.\  {{\bar \omega }_{m,n}} = \left\{ {{e^{j\psi }},\psi  \in \Psi } \right\},|{w_m}| = 1.
	\end{aligned}
\end{equation}

Although the solution to this problem may not be globally optimal due to the nonlinear mapping of the steering vectors, a small-scale search on this basis can be very close to the optimal value. For (\ref{eq63}), it is divided into two problems. For the fixed ${\rm w_m}$, the optimal discrete phase shift can be solved by (\ref{eq62}). Then the values of ${\rm w_m}$ are searched under fixed ${\bar \omega }_{m,n}$. The objective function is equal to $\sum\limits_{n = 1}^N {|(1 + {e^{j({p_m} + \bar \psi _{m,n}^{} - \psi _{m,n}^{opt})}})|} $, where ${p_m}$ is the phase of the weighted value. Therefore, the problem is transformed into
\begin{equation}\label{eq64}
	\begin{aligned}
	&	\mathop {\min }\limits_{{p_m}} \sum\limits_{n = 1}^N {|{p_m} + \bar \psi _{m,n}^{} - \psi _{m,n}^{opt}|}, \\
&s.t. \ {p_m} \in [ - \pi/2^B ,\pi/2^B ].
	\end{aligned}
\end{equation}	

The problem is equivalent to finding the shortest sum of distances to all points, and has a closed-form solution with
\begin{equation}\label{eq65}
	\begin{aligned}
 {p_m} = \frac{1}{N}{{\sum\limits_{n = 1}^N {\psi _{m,n}^{opt} - \bar \psi _{m,n}^{}} }}.
	\end{aligned}
\end{equation}	

The sub-optimal discrete phase shifts and weighted values are obtained by repeating this process. It can be demonstrated that the gap between the discrete phase shifts and the optimal continuous phase shifts only decreases during each iteration.

\subsection{Overall algorithm}

The  alternating optimization algorithm  is given in Algorithm \ref{alg2}. During the iteration, the target SINR is updated by the bisection search. The lower bound of the search scope is set as $0$ and the upper bound is set as
\begin{equation}\label{eq66}
	\begin{aligned}
{\gamma _{\max }} &= \frac{{{P_a}\Vert{{\bf{w}}^H}{\Vert^2} \Vert{\bf{\Phi }}({{{\bf{\bar \theta }}}_a}) \odot {\bf{A}}({{{\bf{\bar \theta }}}_a}){\Vert^2}\Vert{{{\bf{\bar g}}}_{ab}}{\Vert^2}}}{{\sigma _n^2}} \\
&= \frac{{{P_a}{M^2}{N^2}{L_a}^2}}{{\sigma _n^2}}.
	\end{aligned}
\end{equation}	

In each iteration, the target SINR is set as $\gamma= (\gamma_{min}+{\gamma_{max}})/2$. In the inner layer optimization, the optimal continuous phase shifts are obtained by solving problem ${\cal P}_{1.2}$. If a feasible solution is found, the search range is extended to the range between the current SINR and its maximum value. The process is then repeated to determine the maximum feasible SINR value. In the outer layer, given the optimal continuous phase shifts, the optimal weighted values of antennas and discrete phase shifts of DMA elements are obtained by iterative updating. For the computational complexity of the proposed robust algorithm, the complexity mainly comes from the SDP problems in (\ref{eq62}). According to \cite{COMPA1}, the complexity for solving (\ref{eq62}) is given by ${\cal O}({\tilde V_1^{3.5}}log{\frac{\gamma_{max}}{\kappa}})$ where $\tilde V_1=N^2M^2+2M+4$ is the number of variables to be optimized.

	\begin{algorithm}[htbp]
	\caption{Proposed algorithm for solving problem ${\cal P}_1$}
	\label{alg2}
	{Initialization:  Set the maximum number of iterations $L_1$, the tolerance thresholds $\kappa$ and $\epsilon$,and the searching range with $\gamma_{min}=0$ and $\gamma_{max}$ in (\ref{eq66}). Initialize weighted value ${\rm w}_0$, $l_1=1$ and $E^{(0)}=0$. Obtain the estimation of DoA and CSI. }
	{\begin{algorithmic}[1]
			\While {  $ \gamma_{max}- \gamma_{min} > \kappa $}
			\State  $\gamma_{opt}= (\gamma_{min}+{\gamma_{max}})/2$.
			\State Substitute $\gamma_{opt}$  into (\ref{eq62});
			\If{(\ref{eq62}) is feasiable}
			\State $\quad \gamma_{min}=\gamma_{opt}$, and store the optimal $\bf V$;
			\Else
			\State $\quad\gamma_{max}=\gamma_{opt}$;
			\EndIf
			\EndWhile
			\State  $\bf v$ is derived by the eigenvalue decomposition of $\bf V$;
			\State Substitute ${\bf w}^{(0)}$ and $\bf v$ into (\ref{eq64}) and obtain ${\boldsymbol \omega}^{({1})}$ and the sum of the distance ${\bar E}^{(1)}$;
			\While  {${\bar E}^{(l_1)}-{\bar E}^{(l_1-1)}>\epsilon$ and $l_1 <L_1$}
			\State $l_1=l_1+1$;
			\State calculate ${\bf w}^{(l_1)}$ with (\ref{eq65});
			\State Substitute ${\bf w}^{(l_1)}$ and $\bf v$ into (\ref{eq64}) and obtain ${\boldsymbol \omega}^{(l_1)}$ and the sum of the distance ${\bar E}^{(l_1)}$;
			\EndWhile
			\State {\textbf{Output:} } ${\bf w}^{opt}={\bf w}^{(l_1)}$ and ${\boldsymbol \omega}^{opt}={\boldsymbol \omega}^{(l_1)}$.
	\end{algorithmic}}
\end{algorithm}

\section{Simulation results}

In this section, we present numerical results to evaluate the advantage of the RHA in terms of the transmission performance and robustness in the finite multipath scattering environment. 

\subsection{ Simulation Setting}

Unless otherwise specified, the simulation parameters are set as follows: The receiver receives 4 multipaths from both the transmitter and the jammer, with their respective DoAs randomly distributed in the range of [-90, 90] degrees relative to the normal line of the array. The noise power is set to -90dBm. For simplicity without loss of generality, we ignore the fading of individual multipaths and model them uniformly as the multipaths' power arriving at the receiver, i.e. the power of the signal paths and the jamming paths are 20dB and 40dB to the noise power, respectively. Following the settings in \cite{DMA3}, each heterogeneous antenna in the RHA contains 8 elements spaced at $\lambda/4$, and the array consists of 4 antennas spaced at $2.5\lambda$. For the DMA elements, each element is controlled by 3 bits with 8 states, which are uniformly distributed in the phase shift range. The waveguide parameters are set to $\alpha_t=0.1$ and $\beta_t=2\pi n_gf/c$ with guide index $n_g=2.5$. The simulation results are obtained by 5000 Monte Carlo simulations.

\begin{figure}[t]
\centering
\includegraphics[scale=0.55]{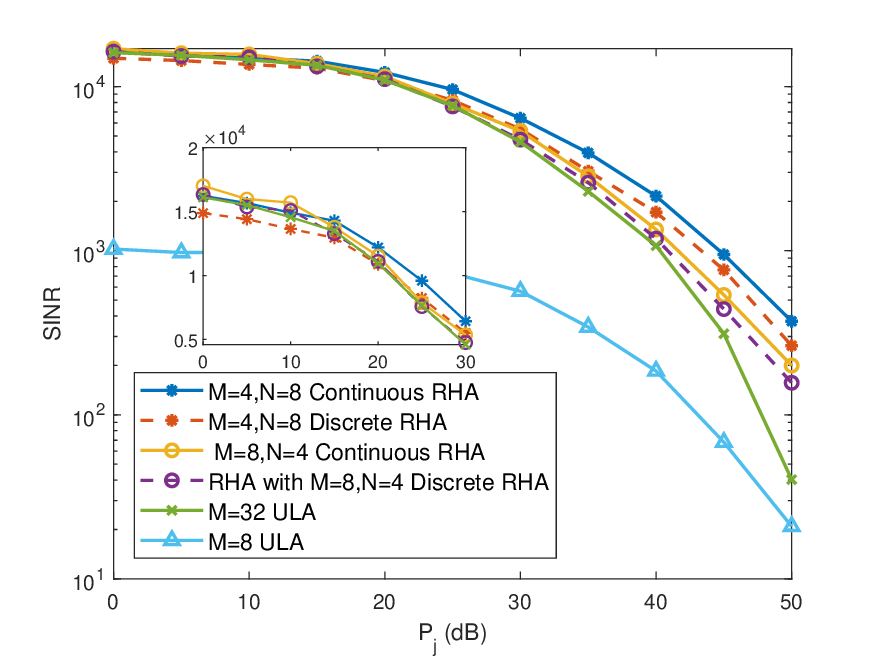}
\vspace{-0.3cm}
\caption{Receiving SINR versus jamming power}
\label{fig5}
\vspace{-0.5cm}
\end{figure}

\begin{figure}[t]
	\centering
	\includegraphics[scale=0.55]{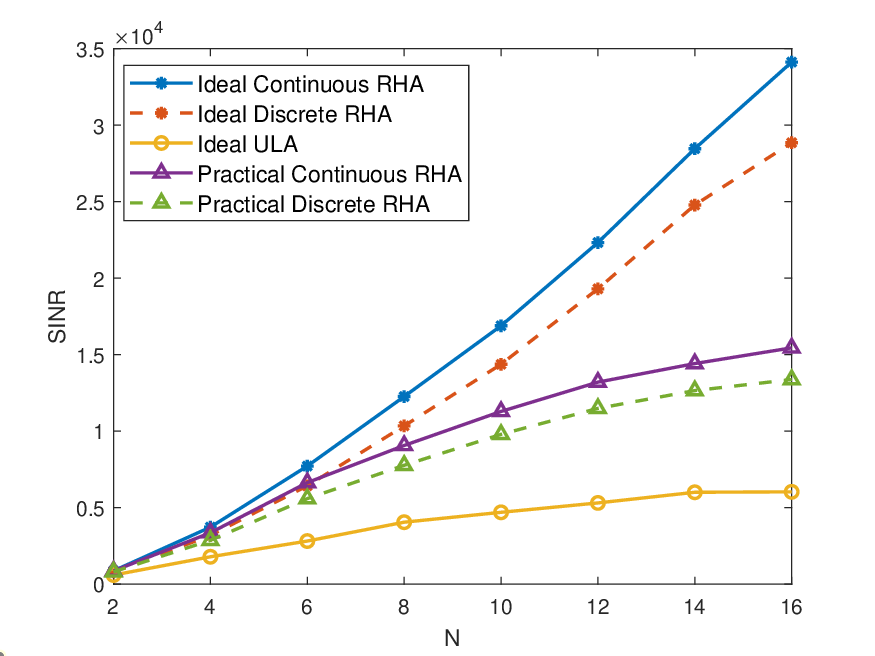}
	\vspace{-0.3cm}
	\caption{Receiving SINR versus the number of elements $N$}
	\label{fig7}
	\vspace{-0.5cm}
\end{figure}

\subsection{ The Influence of Different System Parameters}
To compare the capability in jamming immunity between RHA and ULA, we first eliminate the effect of estimation errors and focus on the influence of the array architecture and system parameters on the SINR. Fig.\ref{fig5} compares the SINR of different array structure without estimation errors under different jamming power.  It can be observed that the proposed RHA scheme consistently outperforms a 32-element ULA with the same array aperture in the continuous case and significantly exceeds an 8-element ULA with the same antenna gain. In the discrete case, the SINR of the RHA with $M=8, N=4$ is always higher than that of a 32-element ULA, demonstrating the effectiveness of the proposed RHA.
In particular, when the jamming power is high, the SINR of a ULA decreases rapidly while the proposed RHA  maintains a high SINR. This is because the RHA can process signals before the RF link to ensure that signals are not submerged in the ADC sampling, whereas conventional homogeneous arrays cannot. At the same time, it can be observed that the RHA with $M=8, N=4$ has a higher SINR when the jamming power is low. However, the RHA with $M=4, N=8$ performs better when the jamming power is high. This also indicates that different RHA configuration modes must be designed according to the environmental conditions to better eliminate jamming.

\begin{figure}
	\centering
	\includegraphics[scale=0.55]{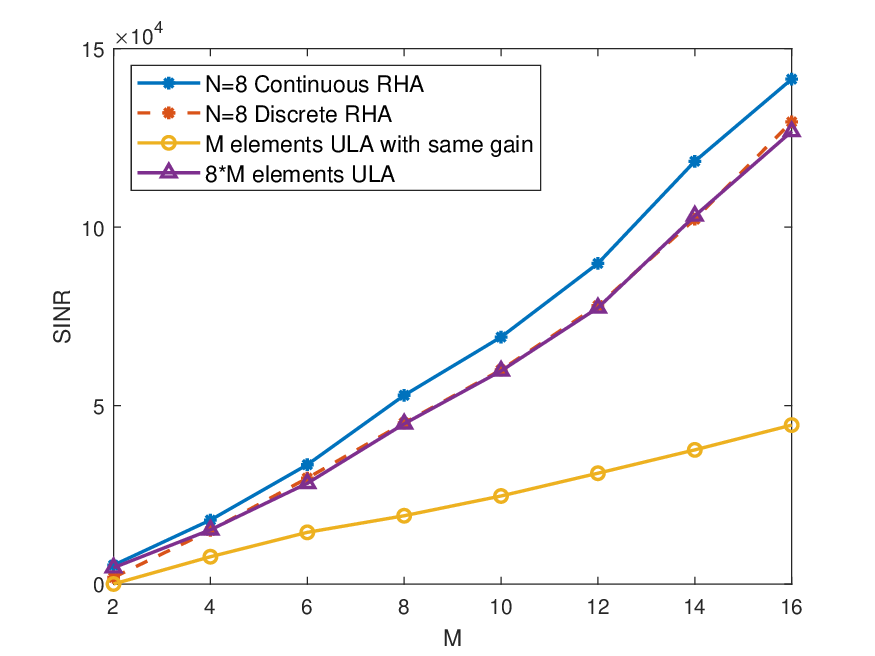}
	\vspace{-0.3cm}
	\caption{ Receiving SINR versus the number of array's antenna $M$ }
	\label{fig9}
	\vspace{-0.5cm}
\end{figure}
\begin{figure}[t]
	\centering
	\includegraphics[scale=0.55]{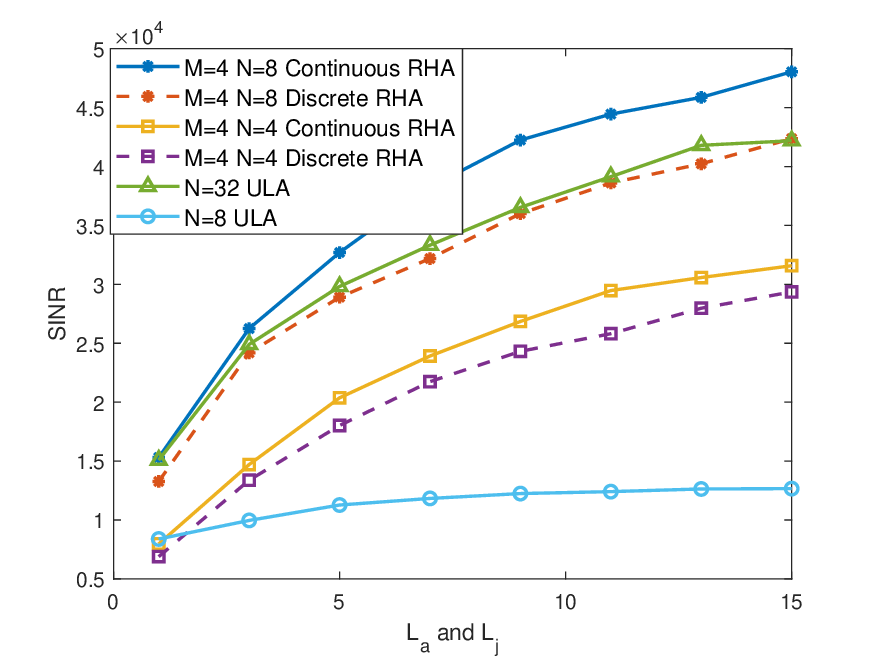}
	\vspace{-0.3cm}
	\caption{Receiving SINR versus the number of multipaths}
	\label{fig8}
	\vspace{-0.5cm}
\end{figure}

We simulate the number of elements on each antenna under a fixed aperture. In the ideal scenario, we ignore the actual physical dimensions and mutual coupling effects of the elements. Under these conditions, the SINR of the RHA increases linearly with the number of elements and at a much faster rate than the ULA. However, as the number of elements increases, the gap between discrete and continuous phase shifts also increases. This is due to the inability of discrete phase shifts to achieve sufficient DoF, resulting in a reduced growth rate.
In reality, it is not possible to achieve infinite performance improvement by increasing the number of elements within a finite range due to physical constraints such as electronic components and mutual coupling. Therefore, we consider a more realistic scenario and simulate the coupling between elements. The mutual coupling effect is obtained through CST simulation and modeled as the power of electromagnetic waves emitted from one element to other elements, i.e. S21 parameter. The actual energy that can be radiated outward is calculated using the input energy and the S21 parameter of the element. Under these conditions, the SINR still increases rapidly when the number of elements is small. However, when the number of elements reaches 8, the rate of increase slows significantly and eventually approaches an upper bound.

Fig. \ref{fig9} illustrates the relationship between SINR and the number of antennas in the array. Two baselines are established: a ULA with an equivalent number of antennas (i.e., $8M$ antennas) and a ULA with an equivalent aperture gain (i.e., $M$ antennas with $\sqrt{8}$ amplitude gain). For all algorithms, the SINR increases with the number of arrays. The SINR of the $M=32$ ULA is comparable to that of the discrete RHA but lower than that of the continuous RHA. However, adding array elements to the RHA incurs a lower cost than adding them to a traditional ULA. In other words, a heterogeneous antenna with one RF chain is equivalent to several antennas with RF chains in the ULA. This implies that the energy efficiency and hardware requirements of the RHA are significantly lower than those of a ULA.


Fig.\ref{fig8} illustrates the impact of varying the number of multipaths on the SINR. We set the same number of signal and jamming multipaths and change them simultaneously. The SINR increases as the number of multipaths increases, and the RHA exhibits a more rapid rate of increase compared to the ULA with an equivalent number of array elements. Furthermore, increasing the value of $N$ can effectively enhance the resolution capability for multipaths, thereby improving the SINR. Notably, a discrete RHA with $M=4, N=8$ can achieve an SINR comparable to that of a ULA comprising 32 elements. However, as the number of multipaths further increase, the ULA becomes susceptible to the ADC blocking problem, resulting in a slower increase rate. 

\subsection{Transmission Performance with Estimation error}
Further, we compared the performance of several algorithms in the presence of estimation errors to verify the effectiveness and robustness of the schemes.

\begin{figure*}[htbp]
\centering
\subfigure[Receiving SINR versus jamming power]{\includegraphics[width=0.32\linewidth]{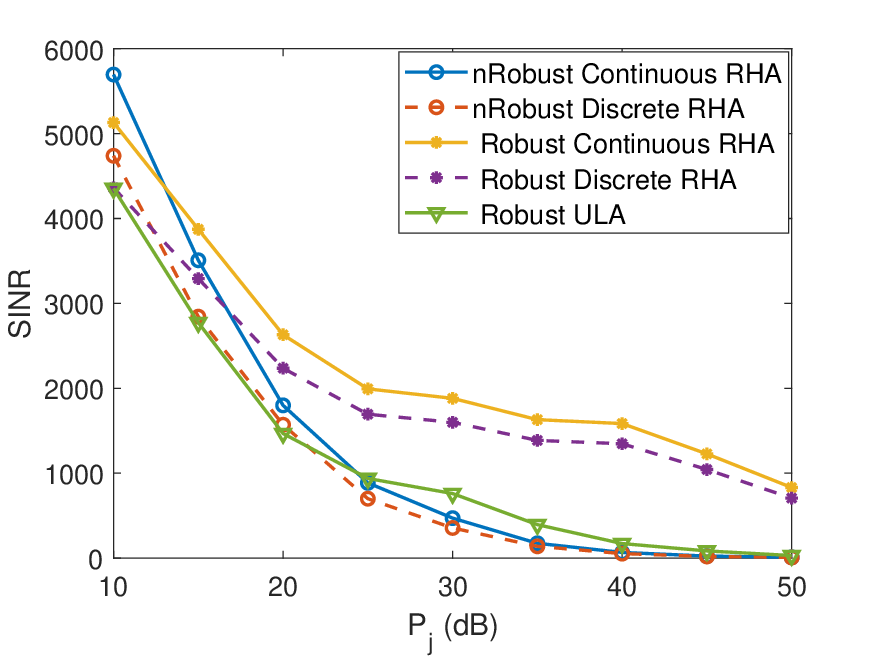}}
\subfigure[Minimum SINR versus jamming power]{\includegraphics[width=0.32\linewidth]{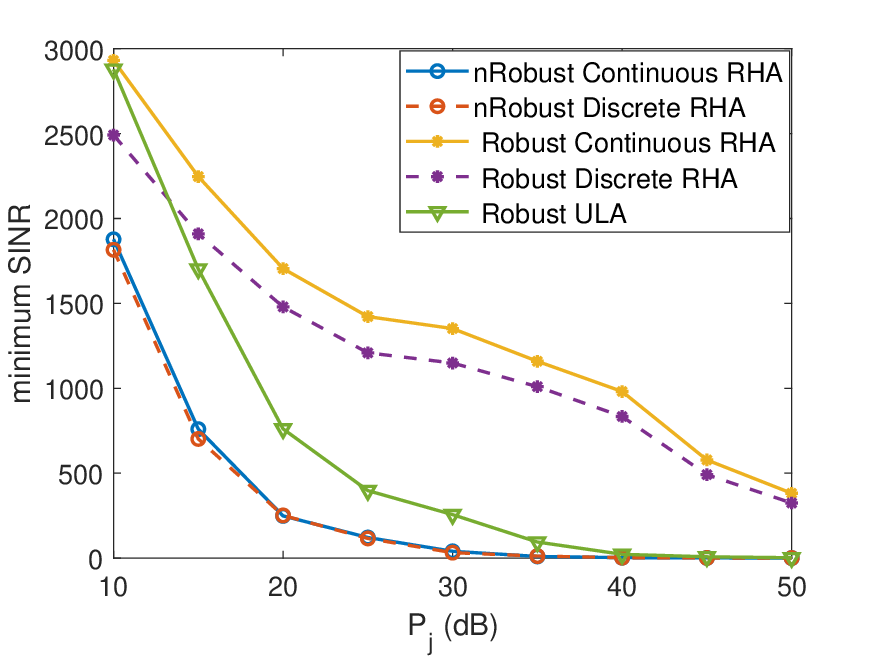}}
\subfigure[Feasible probability versus jamming power]{\includegraphics[width=0.32\linewidth]{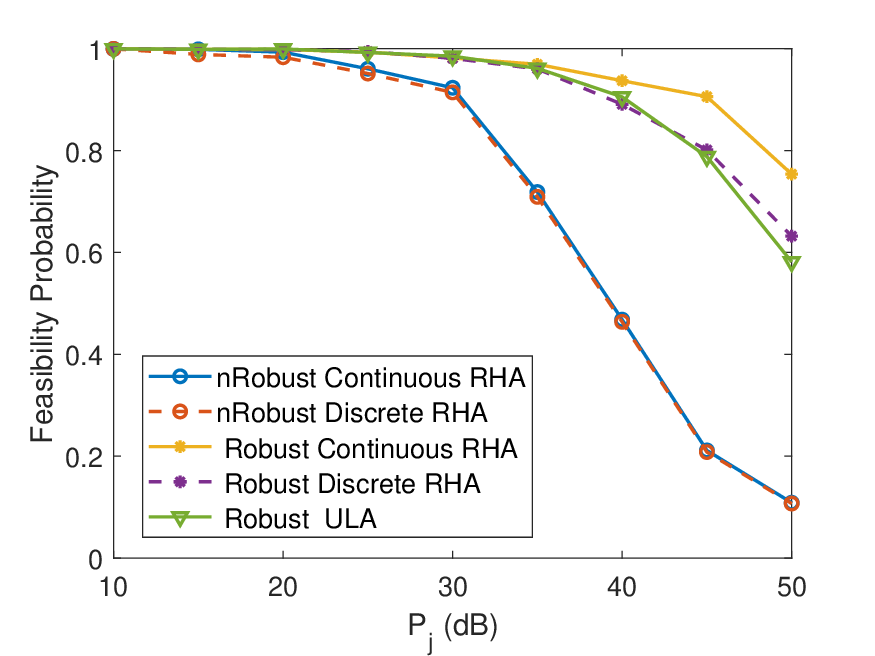}}
\caption{ Receiving SINR and efficiency with different jamming power under imperfect DoA and CSI estimation error}
\label{fig10}
\vspace{-0.5cm}
\end{figure*}

For aligning the effects caused by the errors, we fix the error coefficients as $\rho_{g}=0.1$ and $\rho_{\theta}=0.1$.  Among these algorithms, the nRobust scheme treats the estimation results as the perfect with no estimation error. The compared Robust ULA scheme refers to the robust algorithm presented in \cite{Hou2019}, which only  considers CSI estimation errors.

In Fig.\ref{fig10}(a), the proposed scheme exhibits a higher average SINR compared to the other scheme when $P_j$ is large. As $P_j$ increases, the SINR of nRobust rapidly declines. The Robust ULA scheme also rapidly deteriorates due to its inability to account for the influence of DoA estimation errors. Ultimately, the proposed robust scheme is able to maintain a certain SINR as the jamming power increases, and exhibits a significantly slower downward trend compared to the other two baselines. In Fig.\ref{fig10}(b), we further compare the minimum SINR provided by different schemes. The results indicate that the proposed robust scheme consistently delivers the highest minimum SINR, followed by the Robust ULA scheme. While the nRobust scheme is unable to guarantee a minimum SINR under varying jamming power conditions. Additionally, the minimum SINR achieved by our proposed robust algorithm decreases slowly and is capable of withstanding stronger jamming. Further, we conduct a comparative analysis of the feasibility probability (FP) of several schemes in Fig.\ref{fig10}(c). FP is defined as
\begin{equation}\label{para1}
	\begin{aligned}
		P_F=\sum_{m=1}^{M}Pr\{{\frac{|{{\boldsymbol{\omega }}_m}^H{\bf T}{\boldsymbol{\delta }}({{{\boldsymbol{\bar \theta }}}_a}){{{\bf{\bar g}}}_{ab}}|}{|{{\boldsymbol{\omega }}_m}^H{\bf T}{\boldsymbol{\delta }}({{{\boldsymbol{\bar \theta }}}_j}){{{\bf{\bar g}}}_{jb}}{|^2}}} \ge \xi\},
	\end{aligned}
\end{equation}
which refers to the probability that the solution satisfies the blocking constraints specified in (\ref{eq35}) under 5000 simulations. The results indicate that the proposed scheme exhibits the highest FP, thereby minimizing the occurrence of ADC blocking issues. In contrast, the FP of the non-robust scheme rapidly declines and falls below 50\% when the jamming power exceeds 40 dB, significantly impairing the communication performance. 
\begin{figure*}[htbp]
	\centering
	\subfigure[Receiving SINR versus CSI estimation error ]{\includegraphics[width=0.32\linewidth]{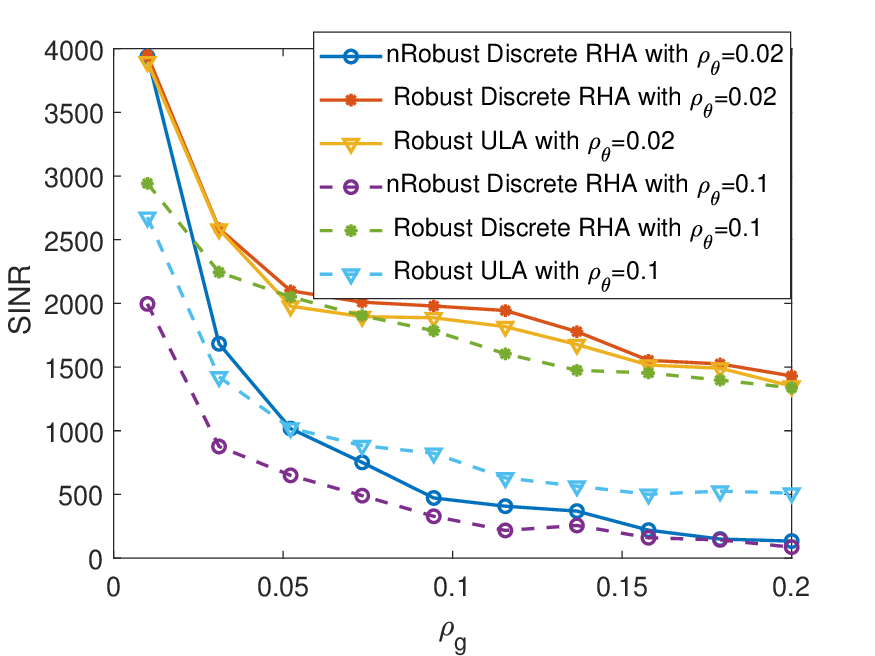}}
	\subfigure[Minimum SINR  versus CSI estimation error]{\includegraphics[width=0.32\linewidth]{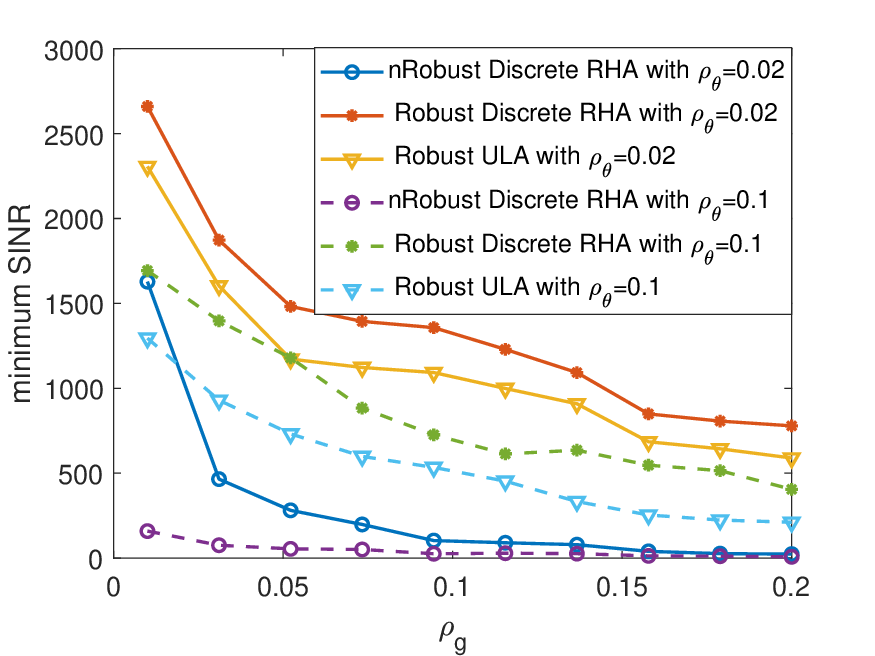}}
	\subfigure[Feasible probability  versus CSI estimation error]{\includegraphics[width=0.32\linewidth]{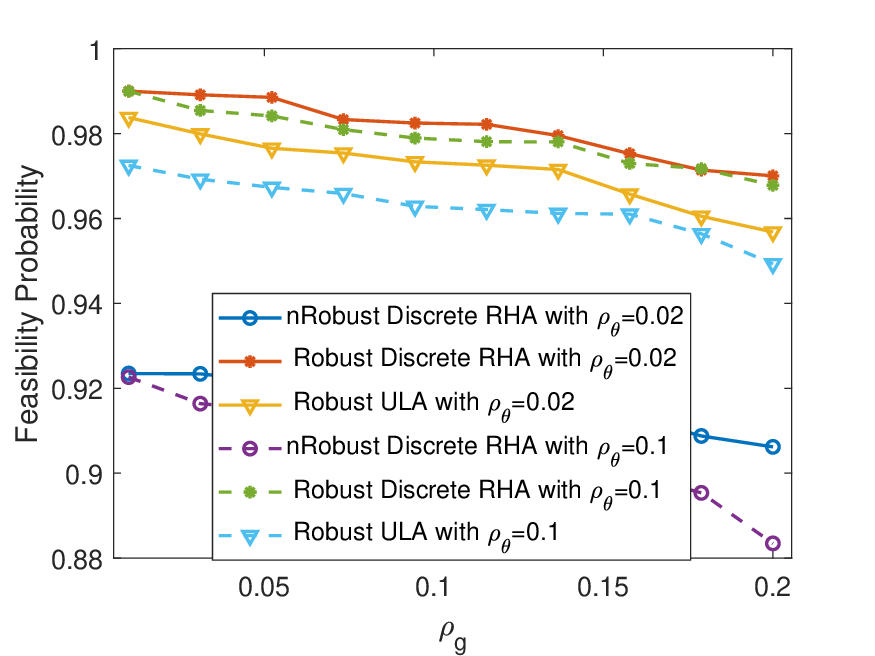}}
	\caption{  Receiving performance under different estimation error  }
	\label{fig12}
	\vspace{-0.5cm}
\end{figure*}

Fig.\ref{fig12}(a) depicts the SINR under varying levels of estimation error. For the DoA estimation error, values of $\rho_{\theta}=0.02$ and $\rho_{\theta}=0.1$ are equal to 1 degree and 5 degrees angle deviation, respectively. When the DoA and CSI estimation errors are small, several schemes exhibit good performance. However, as the CSI estimation error increases, the SINR of the non-robust scheme rapidly declines, while the Robust ULA scheme and the proposed scheme maintain similar SINR levels. An increase in the DoA estimation error results in a weakening of the SINR for all schemes. Nonetheless, the decrease observed in the proposed scheme is the smallest and its SINR ultimately stabilizes at a higher level. The Robust ULA and non-robust schemes do not account for DoA estimation errors, leading to a more pronounced decline in SINR. In Fig.\ref{fig12}(b), we observe that the proposed scheme consistently guarantees a higher minimum SINR. In contrast, other schemes exhibit a rapid decline in minimum SINR as the estimation error increases. With respect to feasible probability in Fig.\ref{fig12}(c), the proposed scheme maintains a feasible probability exceeding 96\% under different DoA and CSI estimation errors. The Robust ULA scheme exhibits a slightly lower feasible probability compared to the proposed scheme, while the non-robust scheme rapidly decreases as the error increases, eventually resulting in a large number of blocking events.
\begin{figure*}[htbp]
	\centering
	\subfigure[Receiving SINR versus jamming power ]{\includegraphics[width=0.32\linewidth]{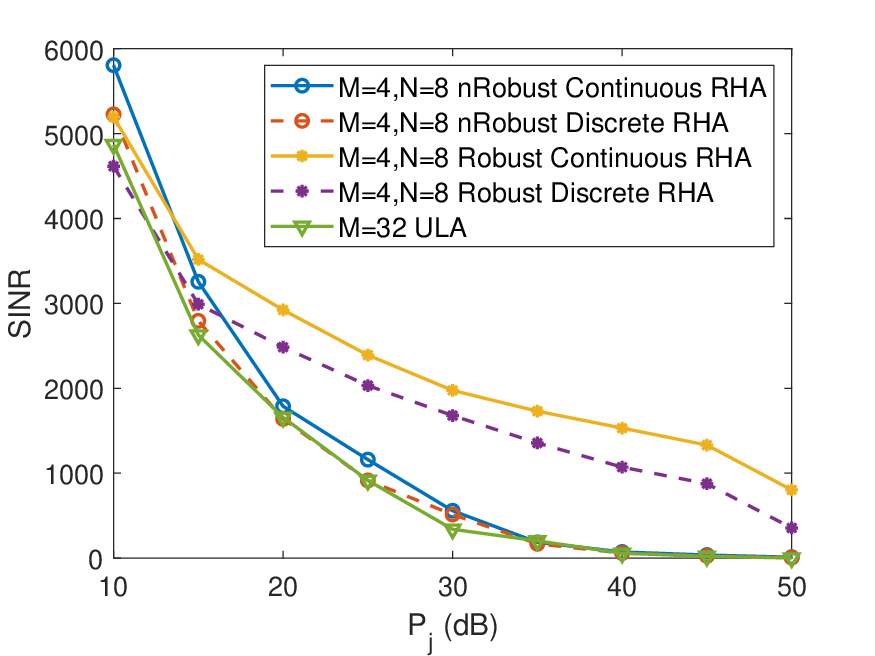}}
	\subfigure[Minimum SINR  versus jamming power]{\includegraphics[width=0.32\linewidth]{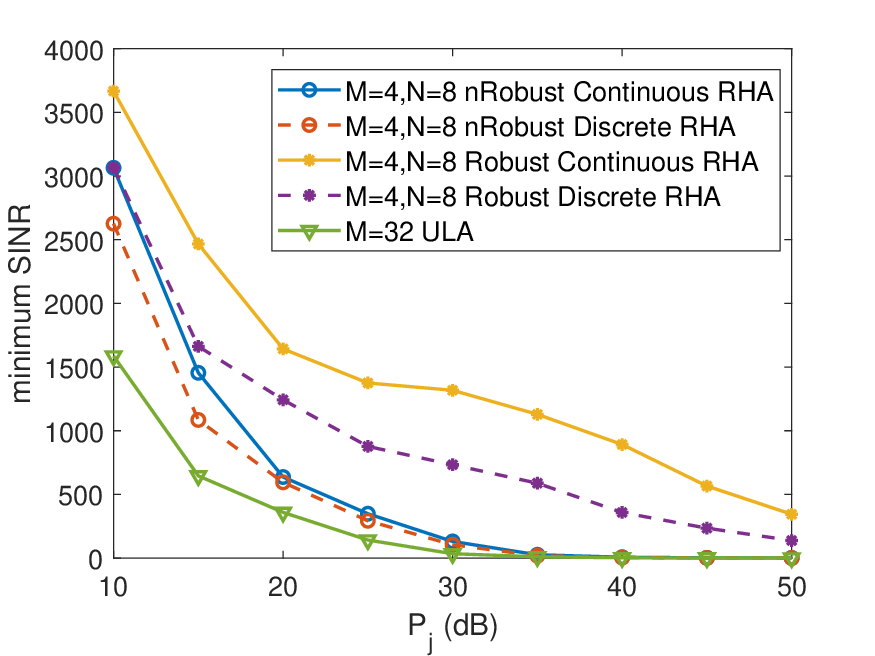}}
	\subfigure[Feasible probability  versus jamming power]{\includegraphics[width=0.32\linewidth]{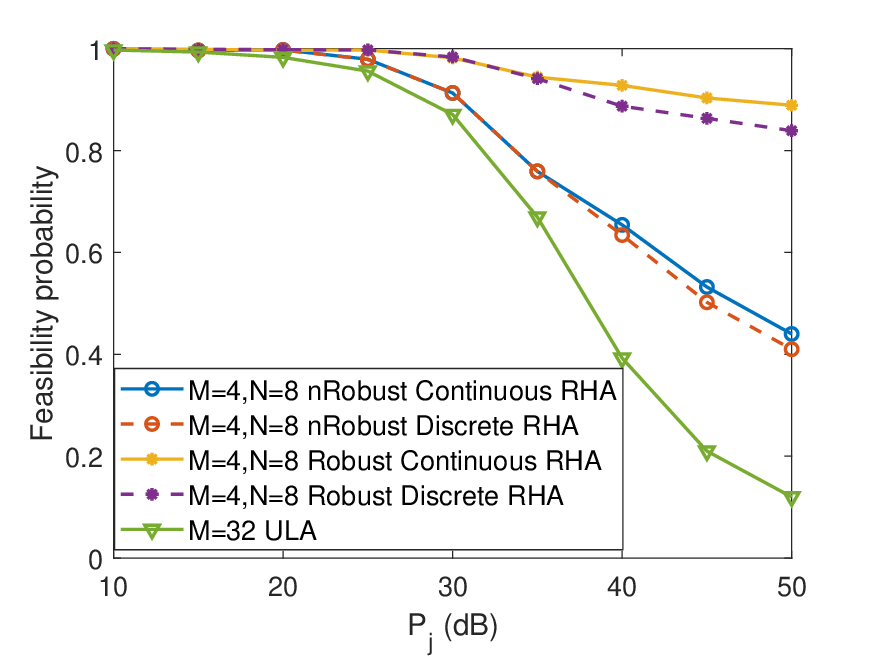}}
	\caption{  System performance under different jamming power }
	\label{fig13}
	\vspace{-0.5cm}
\end{figure*}

Finally, we comprehensively evaluated the performance of the entire system in Fig.\ref{fig13}. We incorporated the DoA and CSI estimation results into algorithms to determine the final weights and phase shifts. The results in Fig.\ref{fig13}(a) indicate that the SINR of the proposed RHA antenna consistently outperforms that of the conventional ULA. This can be attributed to the higher estimation accuracy achieved by the proposed RHA, which in turn yields a greater improvement in SINR. When the jamming and signal powers are similar, the non-robust scheme is capable of achieving a satisfactory SINR. However, as the jamming power increases, the SINR achieved by the robust algorithm remains higher for most jamming power levels. Fig.\ref{fig13}(b) further illustrates that the minimum SINR of the proposed scheme is consistently the highest, while the lower bounds of other schemes rapidly decline and are unable to guarantee a stable SINR. The FPs of various schemes are presented in Fig.\ref{fig13}(c). The results indicate that the FP of ULA rapidly decreases before the jamming power is 30 dB and falls below 50\% after 40 dB, resulting in a pessimistic SINR. In contrast, the non-robust RHA exhibits a higher FP compared to ULA when the jamming power is low. However, it is susceptible to DoA estimation errors, which can also lead to ADC blocking events. The robust scheme demonstrates the highest FP and maintains a probability of over 0.8 of not being blocked after the jamming power is greater than 30 dB.

\section{Conclusion}
In this paper, we investigate a DMA-based RHA architecture to improve anti-jamming capabilities in finite scattering environments. By changing the antenna pattern, RHA provides new gains in estimation and jamming immunity. Specifically, we propose a DOA and CSI estimation method for signal and jamming based on RHA. Then the phase shifts of the DMA elements and the weights of the array elements are jointly designed for maximizing SINR under estimation, and an alternative algorithm is proposed to solve the non-convex optimization problem. Simulation results demonstrate that the proposed RHA schemes improve the received SINR in strong jamming environments, which can be extended to MIMO and hybrid precoding scenarios. Future work will focus on improving the anti-jamming performance of the system through joint transceiver optimization and incorporation of RIS.

\nocite{Win2022}

%

\bibliographystyle{IEEEtran}
\bibliography{IEEEabrv,transusing2}

%
%
%
%
%
%
%
%
%

\end{document}